\documentclass[final,3p,times,twocolumn]{elsarticle}

\usepackage{amssymb}
\usepackage{color}
\usepackage{subfigure}
\usepackage{url}

\usepackage[switch]{lineno}

\usepackage{diagbox}
\usepackage[colorlinks,linkcolor=blue,anchorcolor=blue,citecolor=blue]{hyperref}
\biboptions{sort&compress}
\usepackage{threeparttable}

\journal{NIM A}

\begin{document}

\begin{frontmatter}



\title{Simulation of the 4H-SiC Low Gain Avalanche Diode}

\author[label1,label3]{Tao Yang}
\author[label1,label3]{Yuhang Tan}
\author[label1,label2]{Congcong Wang}

\author[label1,label2]{Xiyuan Zhang\corref{cor1}}
\ead{zhangxiyuan@ihep.ac.cn}

\author[label1,label2]{Xin Shi\corref{cor2}}
\ead{shixin@ihep.ac.cn}

\address[label1]{Institute of High Energy Physics, Chinese Academy of Sciences, 19B Yuquan Road, Shijingshan District, Beijing 100049, China}
\address[label2]{State Key Laboratory of Particle Detection and Electronics, 19B Yuquan Road, Shijingshan District, Beijing 100049, China}
\address[label3]{University of Chinese Academy of Sciences, 19A Yuquan Road, Shijingshan District, Beijing 100049, China}

\cortext[cor1,cor2]{Corresponding authors}

\begin{abstract}

Silicon Carbide device (4H-SiC) has potential radiation hardness, high saturated carrier velocity and low temperature sensitivity theoretically. The Silicon Low Gain Avalanche Diode (LGAD) has been verified to have excellent time performance. Therefore, the 4H-SiC LGAD is introduced in this work for application to detect the Minimum Ionization Particles (MIPs). We provide guidance to determine the thickness and doping level of the gain layer after an analytical analysis. The gain layer thickness $d_{gain}=0.5~\mu m$ is adopted in our design. We design two different types of 4H-SiC LGAD which have two types electric field, and the corresponding leakage current, capacitance and gain are simulated by TCAD tools. Through analysis of the simulation results, the advantages and disadvantages are discussed for two types of 4H-SiC LGAD.

\end{abstract}



\begin{keyword}
LGAD \sep 4H-SiC \sep Impact ionization \sep MIP 


\end{keyword}

\end{frontmatter}



\section{Introduction} \label{sec:introduction}

Searching for high radiation hardness and robustness 4D (time and space) detectors apply to future particle colliders and nuclear reactors where the irradiation flux $>2 \times 10^{16} n_{eq}/cm^{2}$ is \cite{ATLAS} one of the main challenges during the last decade. As the wide band-gap semiconductor material, Silicon Carbide (SiC) is an ideal material to fabricate radiation-resistant devices. Benefited the fast development of the industrial investment of SiC power electronic devices, such as the technology of SiC substrate and fabricating process, design and fabricating the SiC devices to apply in high irradiation flux and high temperature environment become possible. In the major SiC polytypes (3C-SiC, 4H-SiC, 6H-SiC), because the 4H-SiC has the highest band-gap (3.26~eV) corresponding to the highest potential radiation hardness, it is more concerned in the study of high radiation hardness devices.

As the one of significant dimensions for the 4D detectors, time performance strongly depends on carrier velocity and signal noise ratio (S/N)\cite{LGAD_Theory}. The Silicon Low Gain Avalanche Diode (LGAD) has been verified that has better than 50~ps time resolution due to its good S/N between the low gain range (10$\sim$100). And also it has been developed successfully by the various foundries in the past few years\cite{CNM_LGAD,HPK_LGAD,FBK_LGAD,BNL_LGAD,KeweiWU_LGAD,YunyunFAN_NDL,SuyuXIAO_TestBeam,Yuzhen_NDL_33um,YuhanTAN_NDL_CIAE}. However, present studies indicate the collected charges of Silicon LGAD decrease rapidly when the irradiation flux up to $2.5\times10^{15} n_{eq}/cm^{2}$. Besides, the Silicon LGAD needs be cooling to -30~$^{\circ}$C to offset the rapidly rising of leakage current in the extreme irradiation environment\cite{LGAD_RAD_I_G,Ar_Model,YuhanTAN_NDL_CIAE}. Compared with Silicon, 4H-SiC has higher atomic displacement energy, higher saturated velocity and stability in high temperatures. Furthermore, the average e-h pairs generation when  Minimum Ionization Particles (MIP) pass through 1~$\mu m$ 4H-SiC material is 55 $pairs/\mu m$\cite{SiC_MIP_Charges}, which the signal is satisfied with the most of electronics if consider a 50~$\mu m$ 4H-SiC LGAD with 10 gain.

Previous measurements have indicated the 100~ps time resolution could be achieved by 100~$\mu m$ 4H-SiC PIN for detection of MIPs\cite{Tao_NJU_PIN}. The corresponding simulation of 3D 4H-SiC detector indicates the time resolution could reach to 25~ps and shows good stability on high temperature\cite{Yuhan_3D_SiC}. So it has great interest to design a 4H-SiC LGAD for the application of fast particle detection.

Therefore, this work gives a study of 4H-SiC LGAD by simulation. In Sec.~\ref{sec:design}, the thickness and doping level of the gain layer are discussed after analytical analysis. Based on the results in analytical analysis, we design two types of 4H-SiC LGAD  and discuss the simulation results of leakage current, capacitance and gain in Sec.~\ref{sec:simulation}. Finally, the results are summarized in Sec.~\ref{sec:summary}.

\section{4H-SiC LGAD Design} \label{sec:design}


\begin{figure}[htb] 
    \centering
    \includegraphics[scale=0.38]{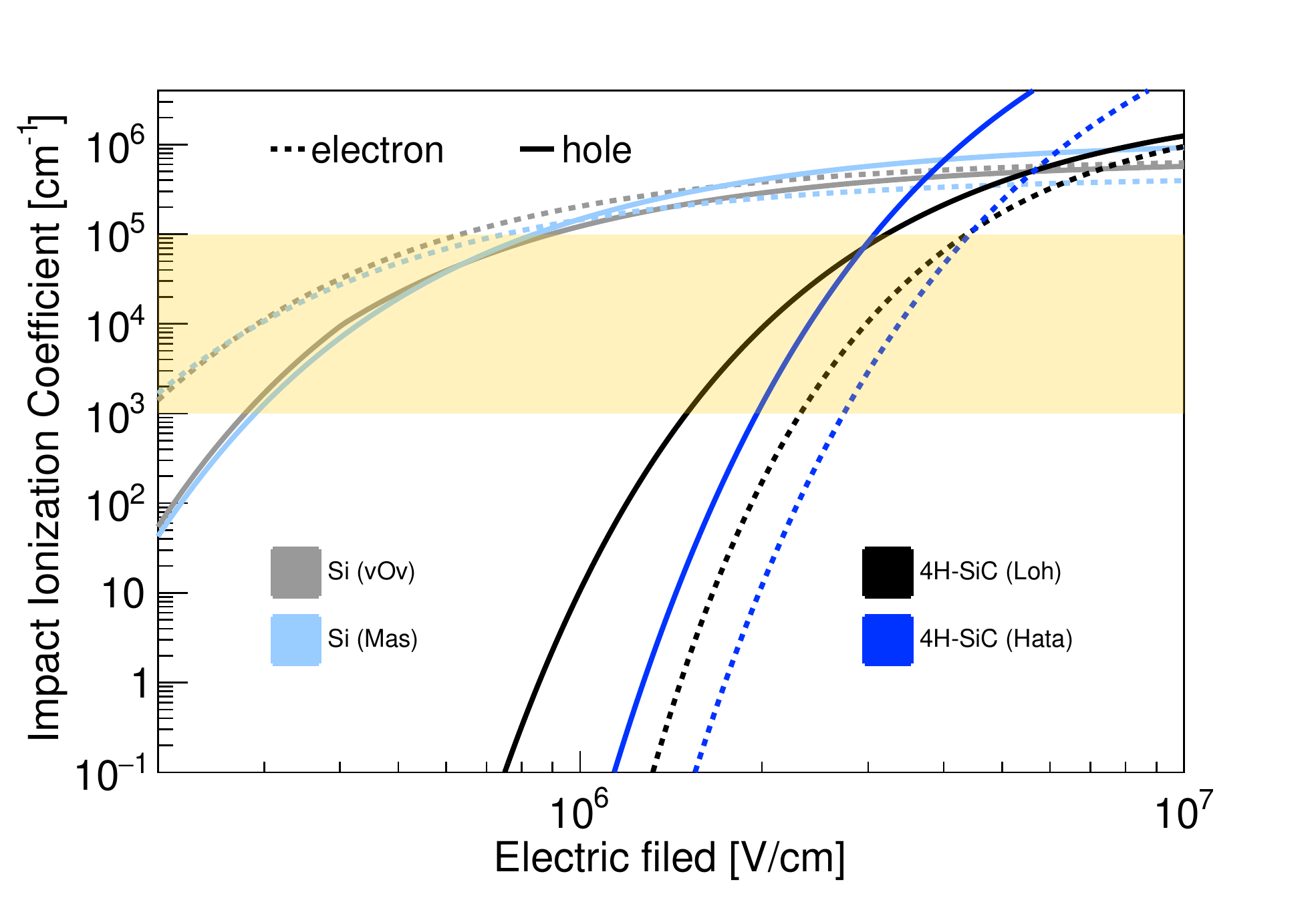}
    \caption{Impact ionization coefficient of Si and 4H-SiC at T = 300~K. The gray lines of Si are from van Overstraeten – de Man model\cite{vOv_Impact_Model}, and cyan lines of Si are from Massey model\cite{Massey_Impact_Model}. They are validated by LGAD simulation in previous studies\cite{LGAD_Rad_Gain_Sim,Tao_LGAD_TCAD}. The balck lines of 4H-SiC are from Loh model\cite{Loh_Impact_Model} which could predicate the breakdown characterization of 4H-SiC accurately in the electric field from 0.9 to 5 MV/cm. The blue lines are from Hatakeyama model on $<0001>$ direction of 4H-SiC which is based on measurement\cite{Hatakeyama_Impact_Model}. The possible low gain region for $\rm 0.1\sim1.0~\mu m$ gain layer is marked by the yellow band.}
    \label{fig:impact_coff}
\end{figure}

To detect the MIPs, the impact ionization coefficient and the generated e-h pairs of 4H-SiC material are essential factors that should be considered in the design of 4H-SiC LGAD. Consider a typical (P++)-(N+)-(N-)-(N++) 4H-SiC LGAD structure, where the P++ layer is the anode, N+ layer is the gain layer with high electric field, N- layer is the bulk layer which the thickness dominates the initial e-h pairs generation, and N++ is cathode layer. We provide some basic policies of 4H-SiC LGAD below after analytical analysis.

The thickness and electric field level of the gain layer is determined depending on the impact ionization coefficient. \figurename{~\ref{fig:impact_coff}} lists the impact ionization coefficients of Si and 4H-SiC from different models. Based on previous study of P-type Si-LGAD, the electric field is between $2\times10^{5}\sim3\times10^{5}~V/cm$ for the thickness of gain layer $\sim1~\mu m$\cite{Tao_LGAD_TCAD}. Similarly, in terms of the gain layer with the thickness of $\rm 0.1\sim1.0~\mu m$, the impact ionization coefficient of the low gain region ranges from $10^{3}~cm^{-1} \sim 10^{5}~cm^{-1}$, as highlighted in the yellow band of \figurename{~\ref{fig:impact_coff}}, the electric field is about $1\times10^{6}\sim4\times10^{6}~V/cm$ for the 4H-SiC LGAD. To acquire the higher gain/E ratio, the N-type 4H-SiC LGAD is adopted in our work where the impact ionization of initial holes dominates the carrier multiplication.

To acquire enough initial e-h pairs, the thickness of the bulk layer should be considered in design of 4H-SiC LGAD. In our design, a 50 $\mu m$ bulk layer is decided which would be ionized $\sim2500$ initial e-h pairs by MIP. After the multiplication of carries with gain$\sim10$, the collected charges is about 4~fC. However, to utilize the total 50 $\mu m$ bulk layer, it should be depleted under an appropriate low voltage. So it demands the doping level of the bulk layer as low as possible. The high quality, low doping and thick 4H-SiC epitaxy are the main challenges of 4H-SiC LGAD. Fortunately, depending on our previous study, the high quality 100 $\mu m$ 4H-SiC epitaxy layer with the doping level $<1\times10^{14}~cm^{-3}$ has been achieved successfully in the PIN device to detect MIPs. Conservatively, the doping level of the bulk layer is determined to $1\times10^{14}~cm^{-3}$ in the 4H-SiC LGAD design.

\begin{figure}[htb] 
    \centering
    \includegraphics[scale=0.38]{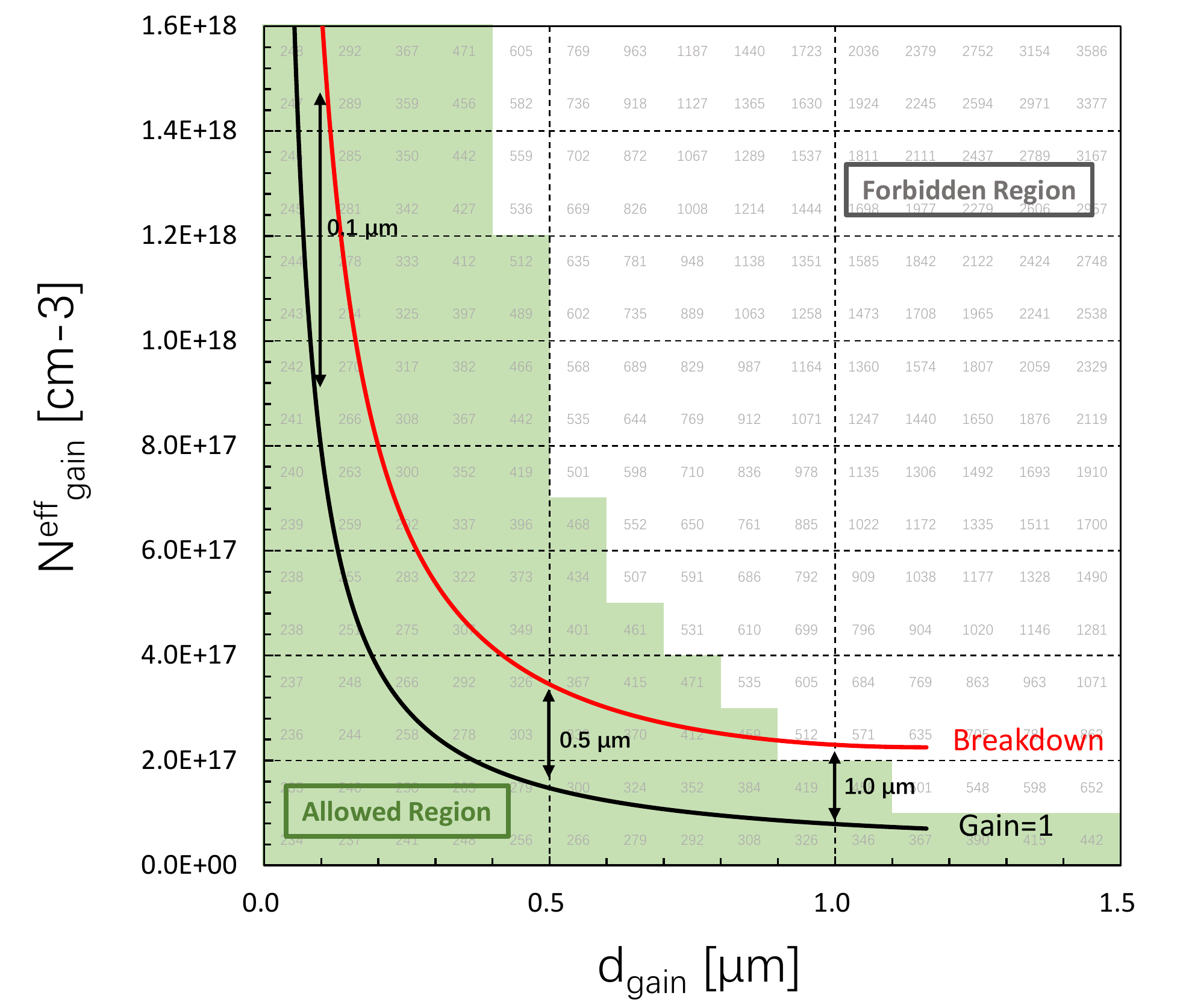}
    \caption{The relation of $N_{gain}^{eff}$ and ${d_{gain}}$ for the (P++)-(N+)-(N-)-(N++) 4H-SiC LGAD structure which has 50 $\mu m$ N- bulk layer with $1\times10^{14}~cm^{-3}$ doping level when U = 500~V. The black line is Gain=1 and red line is Gain=$\infty$ (breakdown). The numbers on the background are the full depletio voltages $V_{FD}~(N_{gain}^{eff},d_{gain})$. The green region is ``allowed region'' which is satisfied with $V_{FD}<V_{BD}$.}
    \label{fig:N_d_relation}
\end{figure}

One of the significant operating conditions of LGAD is

\begin{equation}\label{eq:V_operat}
    V_{FD}<U<V_{BD}
\end{equation}

where $U$ is operating voltage, $V_{FD}$ is full depletion voltage and $V_{BD}$ is breakdown voltage. And the device should have an appropriate gain between $V_{FD}$ and $V_{BD}$. In our design, we demand the gain $\sim 10$ at $U=500V$ and $V_{FD}<\frac{1}{2}V_{BD}$ to guarantee the operating voltage range could be applied by the present source meter like Keithley 2410. Under those limitations, the thickness and doping level of 4H-SiC LGAD could be deduced analytically. 

The $V_{FD}$ is calculated by

\begin{small}
\begin{equation}\label{eq:V_FD}
    V_{FD} =V_{GL}+V_{BULK} = \frac{q \times N_{gain}^{eff} \times d_{gain}^{2} + N_{bulk}^{eff} \times d_{bulk}^{2}}{2 \varepsilon_{SiC} \varepsilon_{0}}
\end{equation}
\end{small}

where q is unit charge; $V_{GL}$ and $V_{BULK}$ are depletion voltage of gain layer and bulk layer; $N_{gain}^{eff}$ and  $N_{bulk}^{eff}$ are doping level; $d_{gain}$ and $d_{bulk}$ are thickness of corresponding layer. The $V_{BULK}$ is estimated $\sim230~V$ if consider $d_{bulk}=50~\mu m$ and $N_{bulk}^{eff}=1\times10^{14}~cm^{-3}$ that mentioned before. Then the the average electric field in gain layer $\bar{E}_{gain}$ could be deduced:

\begin{small}
\begin{equation}\label{eq:E_gain}
    \bar{E}_{gain}=U \times \frac{q \times N_{gain}^{eff} \times d_{gain}^{2}}{q \times N_{gain}^{eff} \times d_{gain}^{2}+q \times N_{bulk}^{eff} \times d_{bulk} ^{2}} \times \frac{1}{d_{g a i n}}
\end{equation}
\end{small}

Consider the average electric field in gain layer is $\bar{E}_{gain}$, the impact ionization coefficient of electron $\alpha_{\mathrm{n}}$ and hole $\alpha_{\mathrm{p}}$ could be calculated by different model. In our work, the Hatakeyama model\cite{Hatakeyama_Impact_Model} is applied because the anisotropy is considered. So the number of e-h pairs $ M(x)$ created by a single of initial e-h pair at position $x$ in the depleted gain layer is\cite{Book_PowerDev}

\begin{equation}\label{eq:Mx}
    M(x)=\frac{\exp \left[\int_{0}^{x}\left(\alpha_{\mathrm{n}}-\alpha_{\mathrm{p}}\right) d x\right]}{1-\int_{0}^{d_{gain}} \alpha_{\mathrm{p}} \exp \left[\int_{0}^{x}\left(\alpha_{\mathrm{n}}-\alpha_{\mathrm{p}}\right) d x\right] d x}
\end{equation}

The breakdown condition is assumed to $\int_{0}^{d_{gain}} \alpha_{\mathrm{p}}=1$ in N-type 4H-SiC LGAD. Consequently, the relation between $N_{gain}^{eff}$ and ${d_{gain}}$ can be determined which shows in \figurename{~\ref{fig:N_d_relation}} depends on Eq.~\ref{eq:V_operat}-\ref{eq:Mx} and other limitations mentioned in our design. The results in \figurename{~\ref{fig:N_d_relation}} indicate the theoretical $N_{gain}^{eff}$ and ${d_{gain}}$ values should be located in ``allowed region'' between the black line (Gain=1) and red line (Gain=$\infty$). Some important information could be extracted in \figurename{~\ref{fig:N_d_relation}} to guid the design of 4H-SiC LGAD: (1)~The ${d_{gain}}$ could not be larger than $1.2~\mu m$ otherwise the device will be premature breakdown before full depletion; (2)~In general,$N_{gain}^{eff}$ decreases with the thickness of ${d_{gain}}$. When ${d_{gain}}$ is less than $0.5~\mu m$, $N_{gain}^{eff}$ decreases rapidly by nearly one order of magnitude with the increase of ${d_{gain}}$ and gradually flattens when ${d_{gain}}$ is greater than 0.5um. Therefore, the ${d_{gain}}<0.5~\mu m$ is not suggested due to high sensitive between the ${d_{gain}}$ and $N_{gain}^{eff}$; (3)~1.0$~\mu m$ thickness is reasonable but easily premature breakdown caused by the fluctuation of $N_{gain}^{eff}$ and ${d_{gain}}$ in practical process. For a comprehensive consideration, the 0.5 $~\mu m$ thickness of gain layer ${d_{gain}}$ is adopted in this work where the doping level $N_{gain}^{eff}$ between $2\times10^{17}~cm^{-3}$ $\sim$ $4\times10^{17}~cm^{-3}$.


\begin{figure*}[!h]
    \centering
    \subfigure[``triangle'' electric field]{ \label{fig:4H-SiC-LGAD-Crossection-Triangle}
    \includegraphics[scale=0.48]{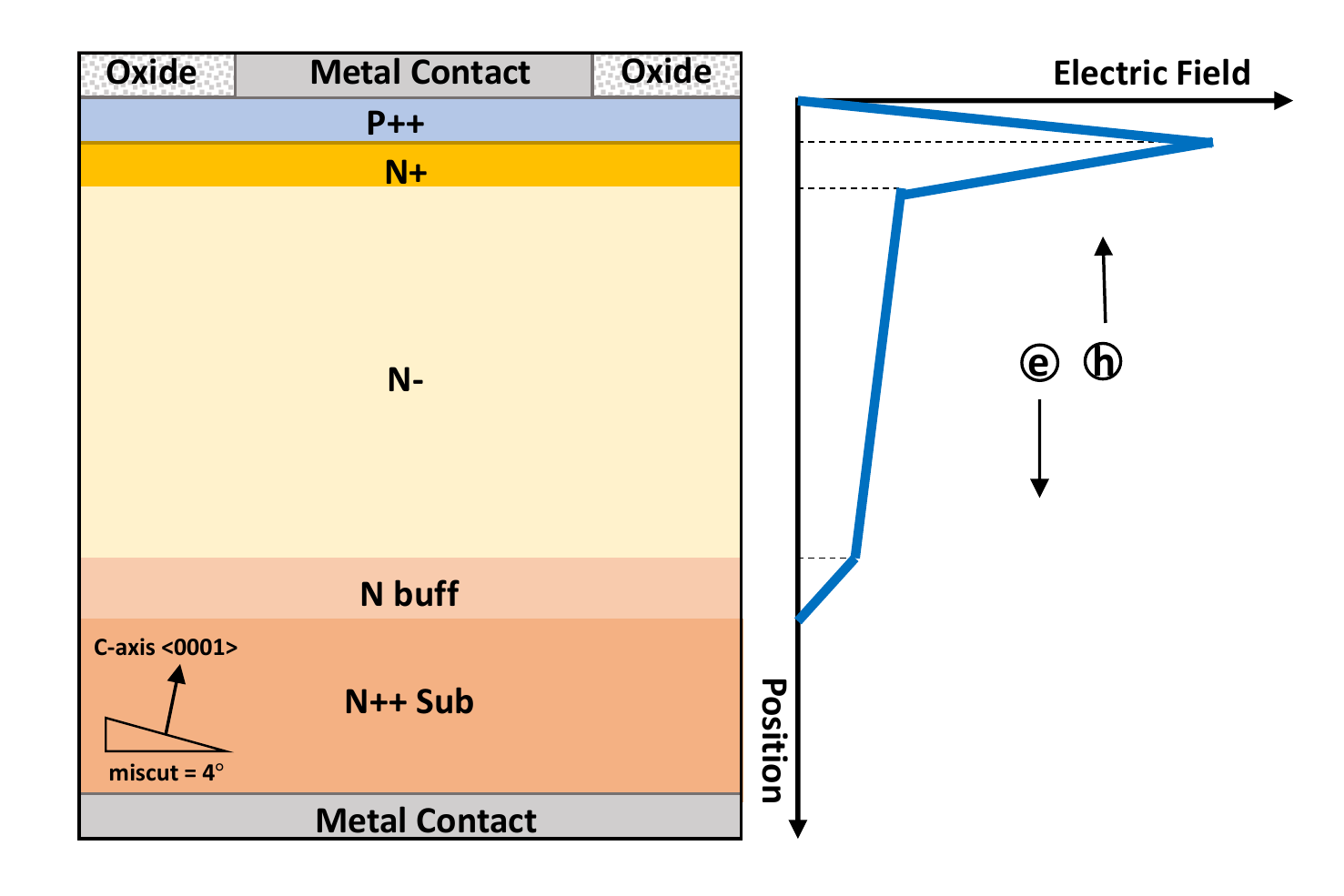}}  
    \subfigure[``trapezoid'' electric field]{ \label{fig:4H-SiC-LGAD-Crossection-Trapezoid}
    \includegraphics[scale=0.48]{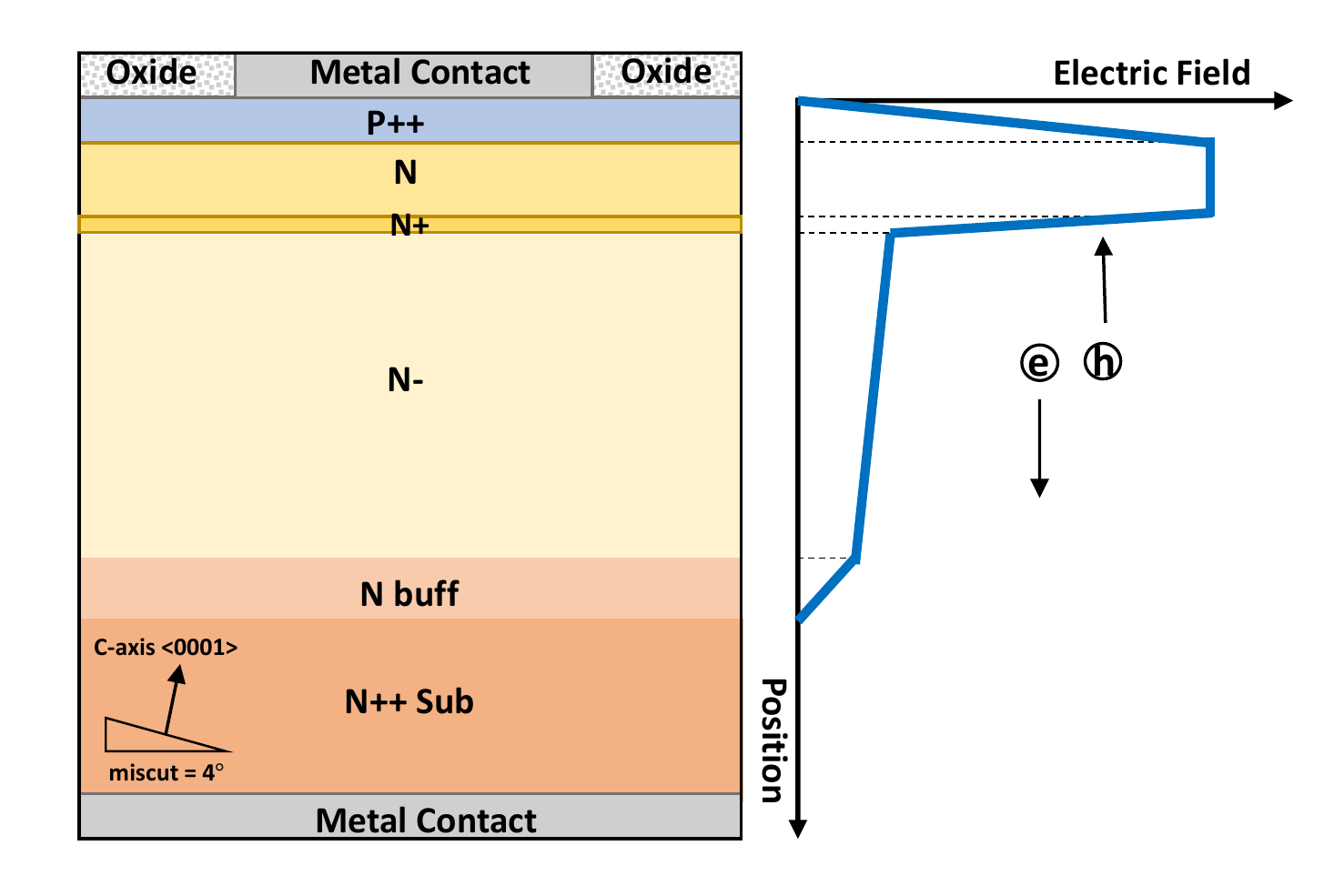}}
    \caption{Longitudinal layers design of the 4H-SiC LGAD studied in this work: (a) ``Triangle'' type with illustration of P+ type ohmic contact layer, N+ gain layer, N- bulk layer, N type buff layer and N++ substrate. (b) ``Trapezoid'' type: with illustration of P+ type ohmic contact layer, N gain layer, N+ electric field control layer,  N- bulk layer, N type buff layer and N++ substrate.}
\end{figure*}

\section{Simulation} \label{sec:simulation}

To verify the results by analytical analysis in Sec.~\ref{sec:design} and determine the specific doping level of gain layer $N_{gain}^{eff}$, we simulate the characteristics of 4H-SiC LGAD with the help of TCAD (Technology Computer Aided Design) tools. 

\subsection{Configuration of physical parameters}

In our simulation, the dielectric constant of 4H-SiC material is $\varepsilon_{SiC}=9.7$. The electron \& hole effective mass and conduction band density-of-state models are from \cite{TCAD_SiCPar_3}. The parameters of the carrier mobility model with doping-dependence refer to \cite{TCAD_SiCPar_1,TCAD_SiCPar_2}. The impact ionization model uses Hatakeyama model \cite{Hatakeyama_Impact_Model}. The MIPs is simulated by heavy ion model of TCAD \cite{TCAD_Manual}. Finally, the wafer system of 4H-SiC has a 4$^{\circ}$ miscut angle in all simulations. 

\subsection{Structure} \label{sec:structure}

Depends on the analytical analysis in Sec.~\ref{sec:design}, a 2D 4H-SiC LGAD structure is designed (see \figurename{~\ref{fig:4H-SiC-LGAD-Crossection-Triangle}}) which has five 4H-SiC layers listed below:

\begin{itemize}
    \item P++ : contact layer, 0.3$~\mu m$, $1\times10^{19}~cm^{-3}$.
    \item N+ : gain layer, the thickness $d_{gain}$  and doping level $N_{gain}^{eff}$ are studied in this work.
    \item N- : bulk layer, 50$~\mu m$, $1\times10^{14}~cm^{-3}$.
    \item N buff : buff layer, 5.0$~\mu m$, $1\times10^{18}~cm^{-3}$.
    \item N++ : substrate layer, 10$~\mu m$, $1\times10^{20}~cm^{-3}$.
\end{itemize}

The width of the structure is 10$~\mu m$ which is satisfied with 2D heavy ion simulation. Because of the ``triangle'' electric field between P++ layer and N+ layer, we address the structure like \figurename{~\ref{fig:4H-SiC-LGAD-Crossection-Triangle}} as Triangle-Type 4H-SiC LGAD. For comparative study, the same structure of $50~ \mu m$ 4H-SiC PIN is designed and the N+ gain layer is removed.

For the same reason, another structure shows in \figurename{~\ref{fig:4H-SiC-LGAD-Crossection-Trapezoid}} is addressed as Trapezoid-Type 4H-SiC LGAD due to the ``trapezoid'' electric field that will be discussed in Sec.~\ref{sec:gain_layer_design}.

\subsection{Simulated I-V and 1/C$^{2}$-V}

\begin{figure*}[!h]
    \centering
    \subfigure[]{ \label{fig:Triangle_Type_IV_0.1um}
    \includegraphics[scale=0.3]{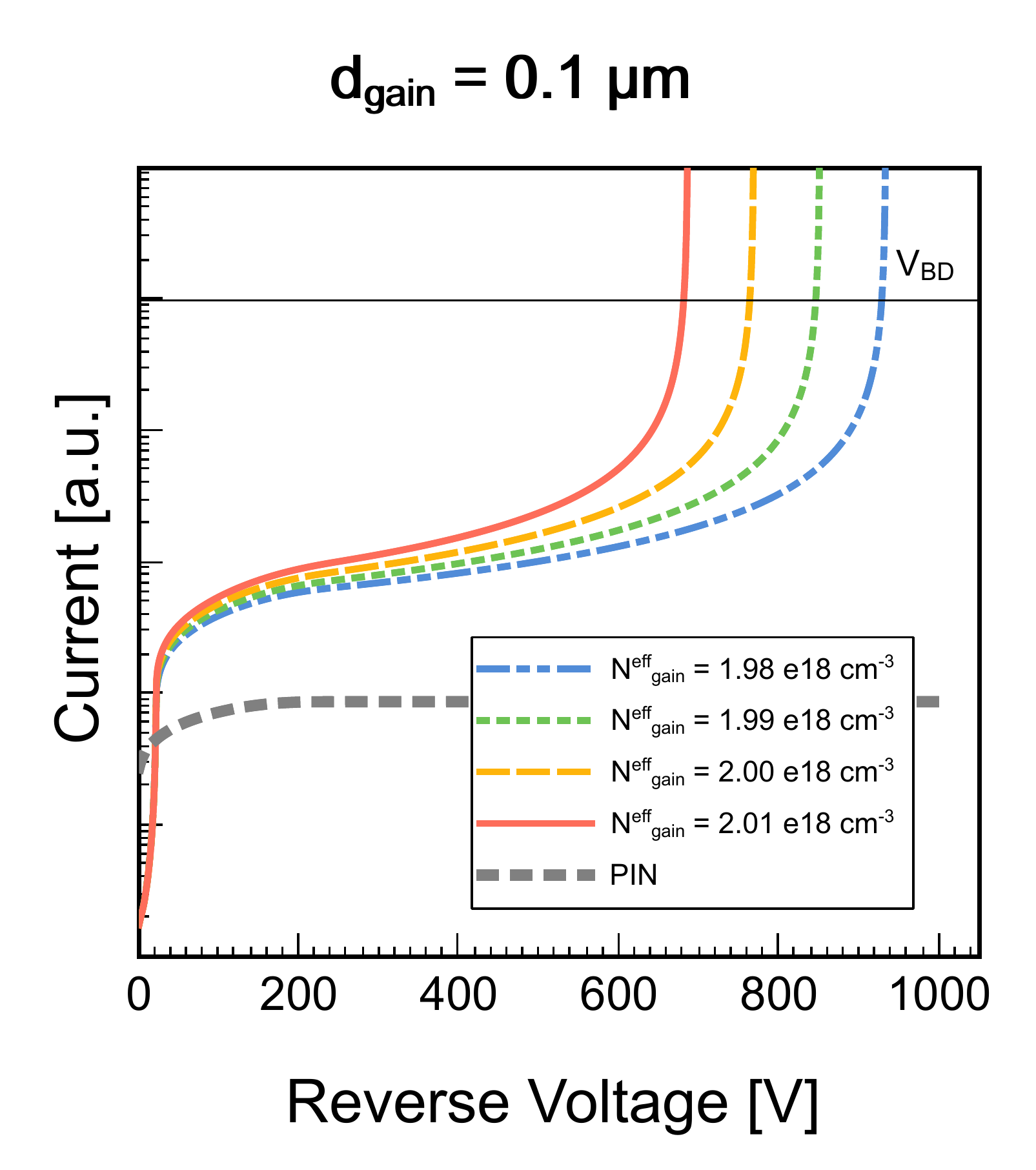}}  
    \subfigure[]{ \label{fig:Triangle_Type_IV_0.5um}
    \includegraphics[scale=0.3]{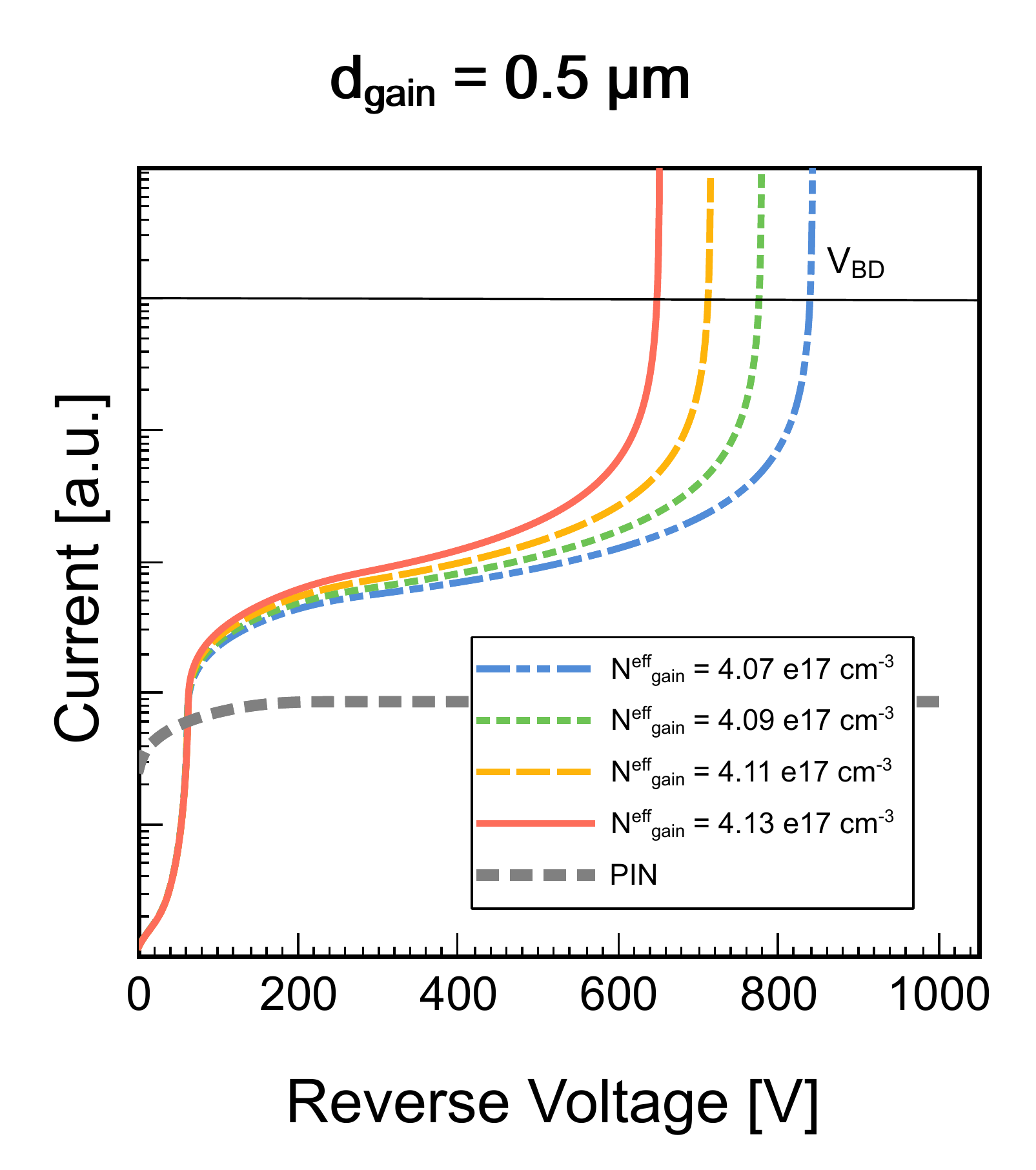}}
    \subfigure[]{ \label{fig:Triangle_Type_IV_1.0um}
    \includegraphics[scale=0.3]{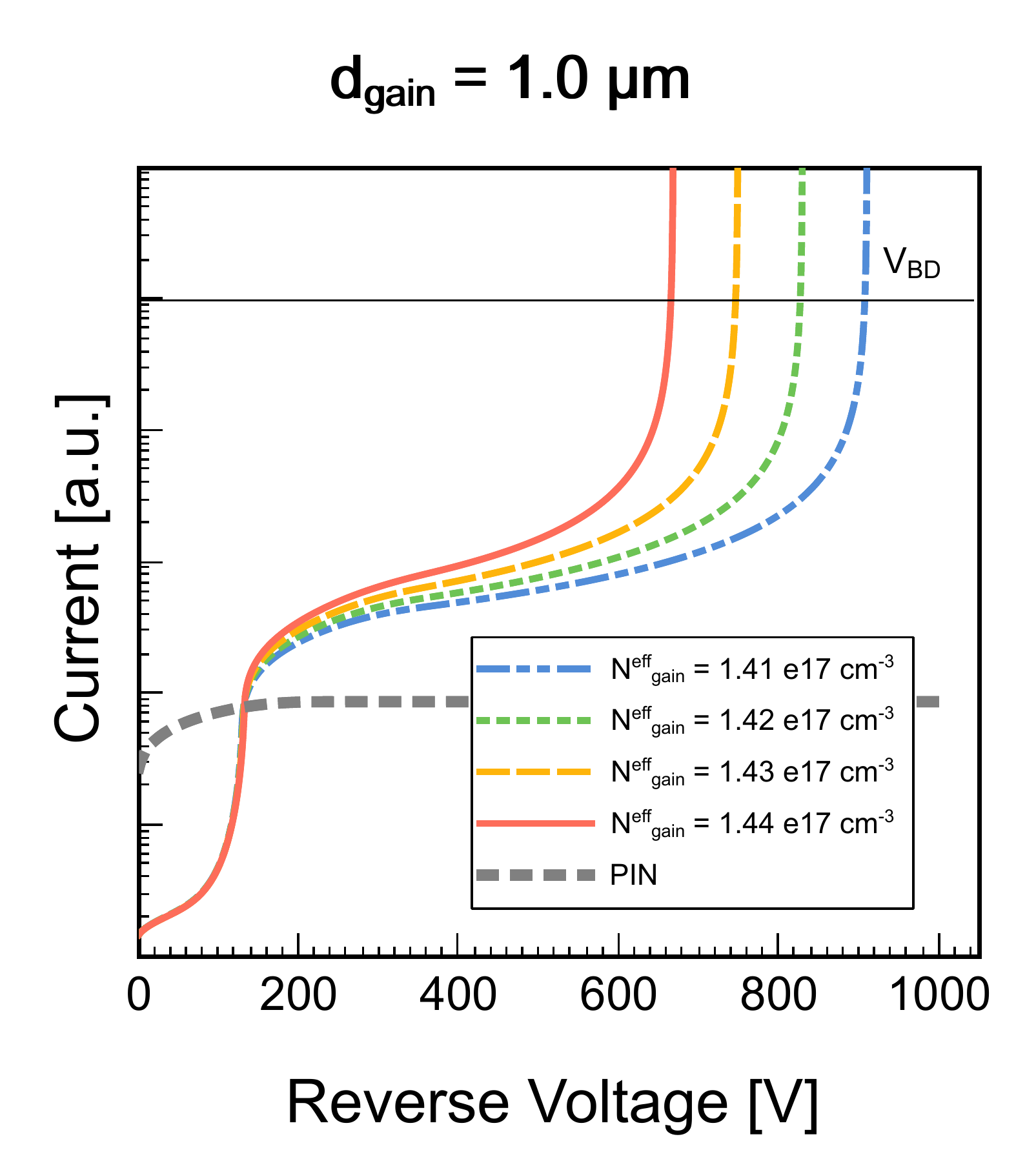}}
    \caption{Leakage current for simulation at 4 different gain layer doping levels of Triangle-Type 4H-SiC LGAD and PIN: (a)~$d_{gain}=0.1~\mu m$; (b)~$d_{gain}=0.5~\mu m$; (c)~$d_{gain}=1.0~\mu m$. The colored lines are Triangle-Type 4H-SiC LGAD. The dotted gray line is PIN.} \label{fig:Triangle_Type_IV}
\end{figure*}

To determine the breakdown voltage $V_{BD}$ for the Triangle-Type 4H-SiC LGAD, the I-V curves are simulated which is shown in \figurename{~\ref{fig:Triangle_Type_IV}}. As \figurename{~\ref{fig:Triangle_Type_IV_0.1um}}$\sim$\figurename{~\ref{fig:Triangle_Type_IV_1.0um}} show, for different gain layer doping levels, the breakdown voltages could be restrained in 500~V$\sim$1000~V in our designs. Under this condition, the order of magnitude of $N_{gain}^{eff}$ is $10^{18}~cm^{-3}$ when $d_{gain}=0.1~\mu m$ and $10^{17}~cm^{-3}$ when $d_{gain}=1.0~\mu m$. It agrees with the predication of analytical analysis in \figurename{~\ref{fig:N_d_relation}}. Compare with the fluctuation interval of $N_{gain}^{eff}$ related with $V_{BD}$ between 500~V$\sim$1000~V, the declining rate of $V_{BD}$ is $\sim 100V ~/~ (10^{16}~cm^{-3})$ for $d_{gain}=0.1~\mu m$, but increases to $\sim 100V ~/~ (10^{15}~cm^{-3})$ for $d_{gain}=1.0~\mu m$. It also agrees with the analysis in Sec.~\ref{sec:design} where the relevance is stronger between $N_{gain}^{eff}$ and ${d_{gain}}$ for thinner $d_{gain}$. It should be noted that the leakage current of 4H-SiC LGAD at low voltages is smaller than PIN is caused by very small volume of depleted region when $U<V_{GL}$. However, the leakage current of all 4H-SiC LGAD designs are larger than PIN after full depletion. The gain of leakage current is $\sim 10$ times indicates the low gain is achieved.

\begin{figure*}[!h]
    \centering
    \subfigure[]{ \label{fig:Triangle_Type_CV_0.1um}
    \includegraphics[scale=0.3]{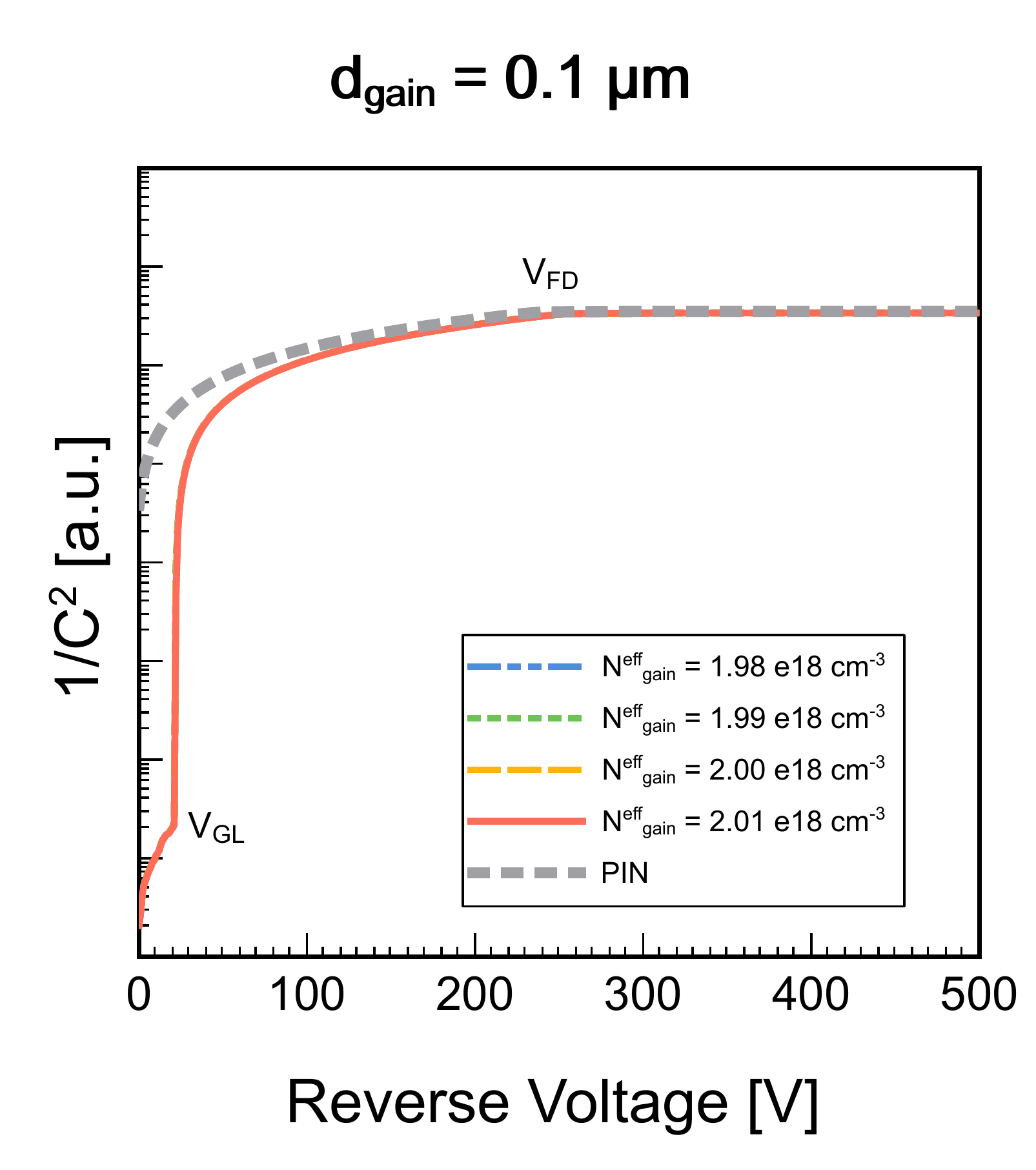}}  
    \subfigure[]{ \label{fig:Triangle_Type_CV_0.5um}
    \includegraphics[scale=0.3]{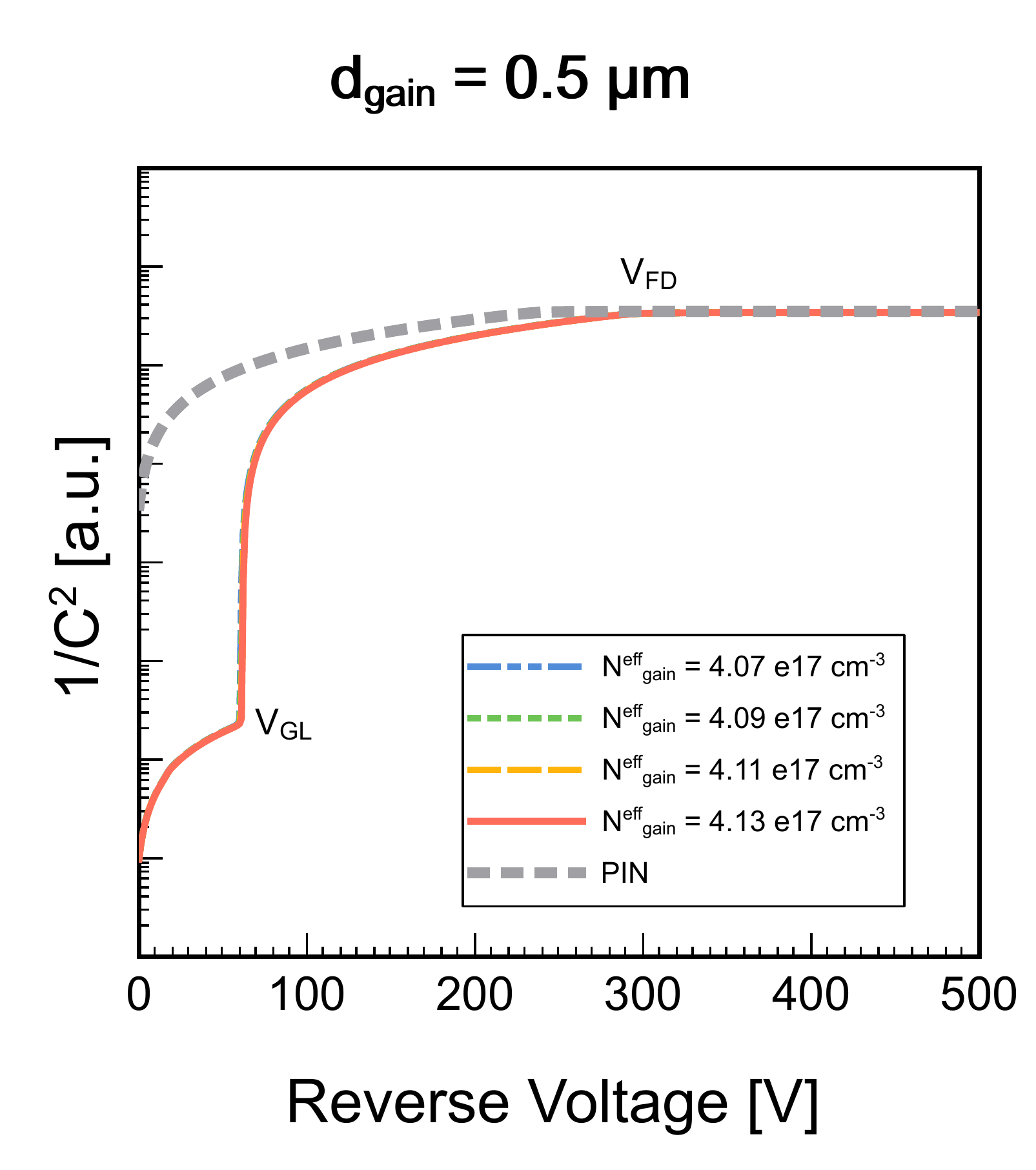}}
    \subfigure[]{ \label{fig:Triangle_Type_CV_1.0um}
    \includegraphics[scale=0.3]{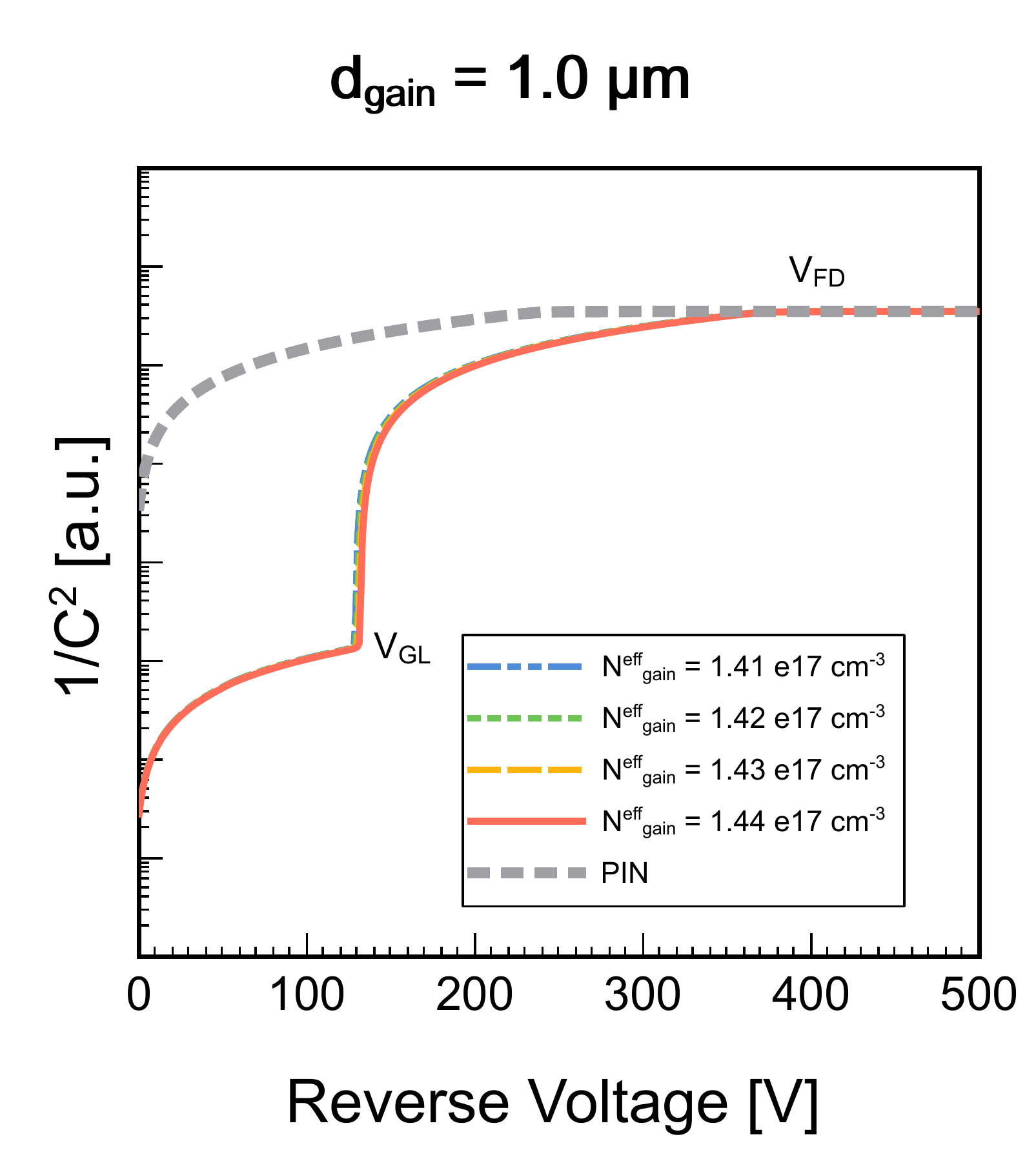}}
    \caption{ 1/C$^{2}$-V for simulation at 4 different gain layer doping levels of Triangle-Type 4H-SiC LGAD and PIN: (a)~$d_{gain}=0.1~\mu m$; (b)~$d_{gain}=0.5~\mu m$; (c)~$d_{gain}=1.0~\mu m$. The colored lines are Triangle-Type 4H-SiC LGAD. The dotted gray line is PIN.} \label{fig:Triangle_Type_CV}
\end{figure*}

To determine the gain layer depletion voltage $V_{GL}$ and full depletion voltage $V_{FD}$ for the Triangle-Type 4H-SiC LGAD, the 1/C$^{2}$-V curves are simulated which is shown in \figurename{~\ref{fig:Triangle_Type_CV}}. As \figurename{~\ref{fig:Triangle_Type_CV_0.1um}}$\sim$\figurename{~\ref{fig:Triangle_Type_CV_1.0um}} show, the influence for fluctuation of $N_{gain}^{eff}$ to $V_{GL}$ could be neglected. For different gain layer doping levels, the $V_{FD}$ of all 4H-SiC LGAD designs are smaller than 500~V that is satisfied with our demand referred in Sec.~\ref{sec:gain_layer_design}. Compare with the fluctuation interval of $N_{gain}^{eff}$ related with $V_{GL}$ for different $d_{gain}$, the $V_{GL}$ increases rapidly from $\sim$140~V for $d_{gain}=0.1~\mu m$ to $\sim$140~V for $d_{gain}=1.0~\mu m$, although the $N_{gain}^{eff}$ has already decreased one order of magnitude. It indicates the thickness of gain layer is leading factor of $V_{GL}$ in 4H-SiC LGAD.

\subsection{Study of gain layer design} \label{sec:gain_layer_design}

\begin{figure}[htb] 
    \centering
    \includegraphics[scale=0.3]{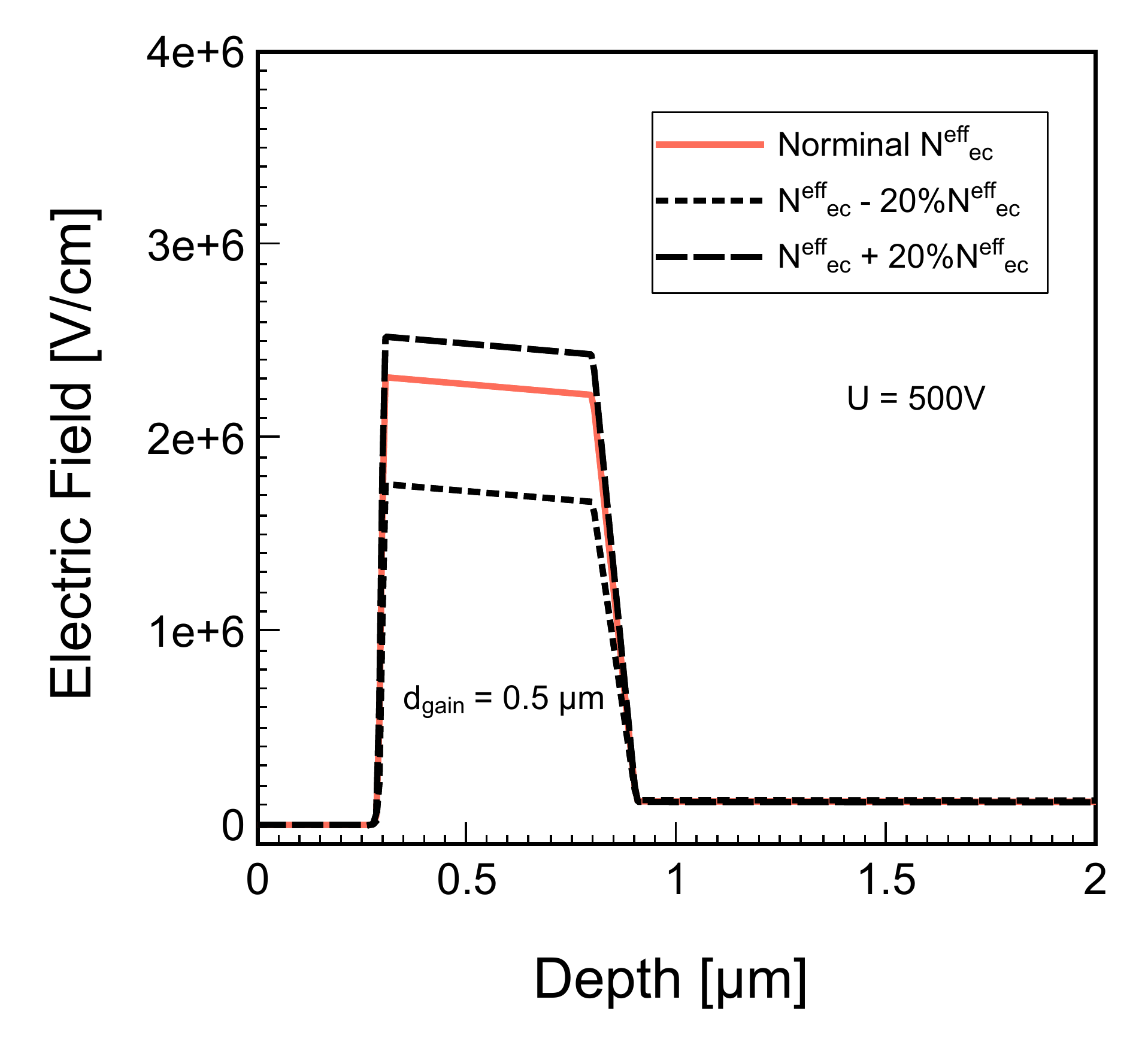}
    \caption{The electronic field in gain layer of Trapezoid-Type 4H-SiC LGAD at U=500~V when when the doping level of electric field control layer $N_{ec}^{eff}$ changed $\pm20$\%.}
    \label{fig:Trapezoid_Type_E_N}
\end{figure}

To control the electric field near the gain layer handily, another 4H-SiC LGAD structure is introduced which has an ``trapezoid'' electric field (See \figurename{~\ref{fig:4H-SiC-LGAD-Crossection-Trapezoid}}). As mentioned in Sec.~\ref{sec:structure}, we address it as Trapezoid-Type 4H-SiC LGAD. Different with the Triangle-Type 4H-SiC LGAD has five layers, the Trapezoid-Type has six layers where another N+ layer called the electric field control layer which is inserted into the interface between the N gain layer and N- bulk layer. The profiles of each layer are listed below:

\begin{itemize}
    \item P++ : contact layer, 0.3$~\mu m$, $1\times10^{19}~cm^{-3}$.
    \item N : gain layer, $1\times10^{16}~cm^{-3}$. The thickness $d_{gain}$ is studied in this work.
    \item N+ : electric field control layer,  0.1$~\mu m$. The doping level $N_{ec}^{eff}$ are studied in this work.
    \item N- : bulk layer, 50$~\mu m$, $1\times10^{14}~cm^{-3}$.
    \item N buff : buff layer, 5.0$~\mu m$, $1\times10^{18}~cm^{-3}$.
    \item N++ : substrate layer, 10$~\mu m$, $1\times10^{20}~cm^{-3}$.
\end{itemize}

In the Trapezoid-Type 4H-SiC LGAD, the width of ``trapezoid'' electric field is dominated by the thickness of gain layer $d_{gain}$ but the electric field level is controlled by the effective doping of electric field control layer $N_{ec}^{eff}$. \figurename{~\ref{fig:Trapezoid_Type_E_N}} demonstrate the control action of electric field when the effective doping of electric field control layer $N_{ec}^{eff}$ changed $\pm20$\%.

\begin{figure}[htb] 
    \centering 
    \subfigure[]{ \label{fig:Trapezoid_Type_IV_0.5um}
    \includegraphics[scale=0.3]{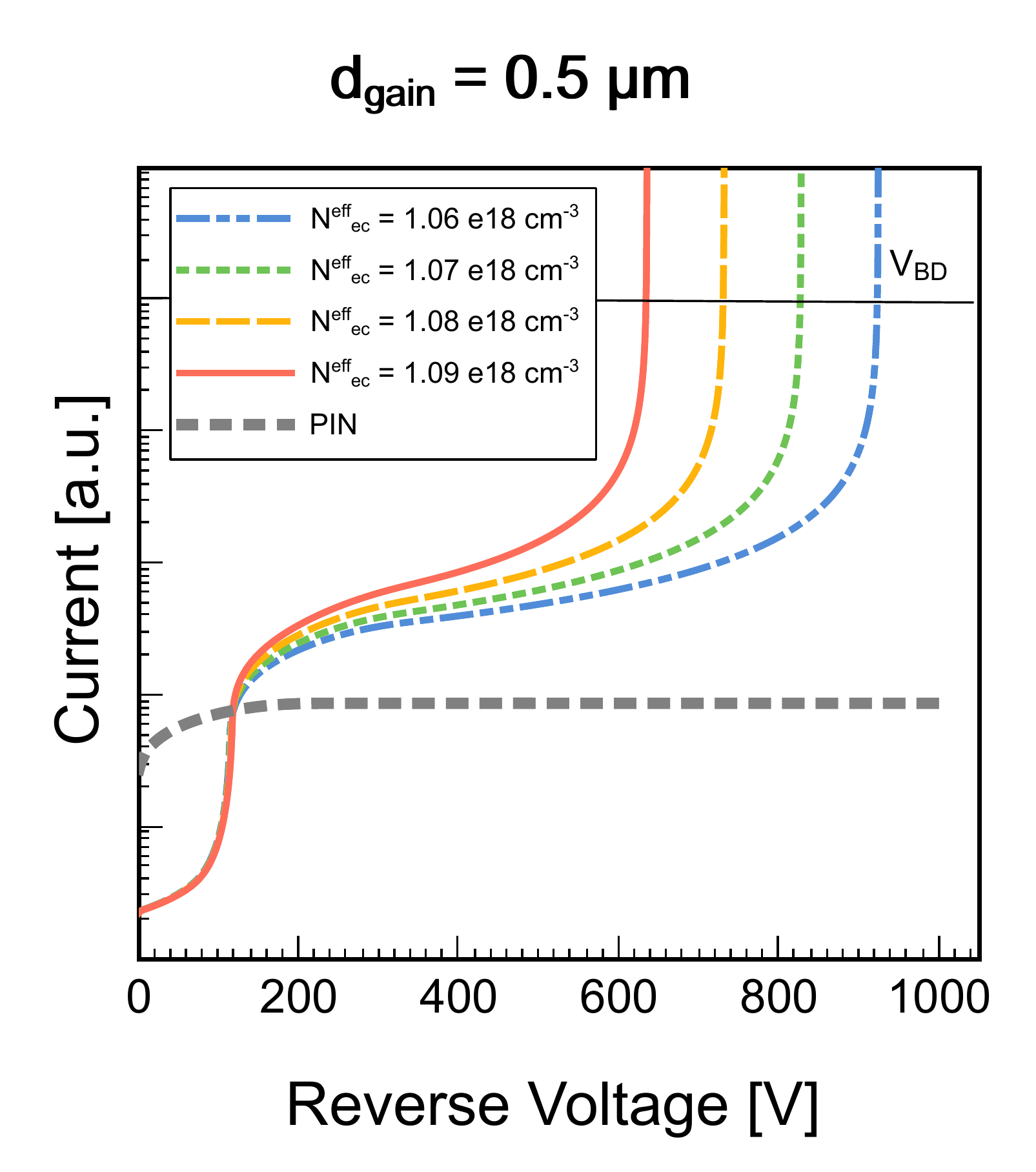}}
    \subfigure[]{ \label{fig:Trapezoid_Type_IV_1.0um}
    \includegraphics[scale=0.3]{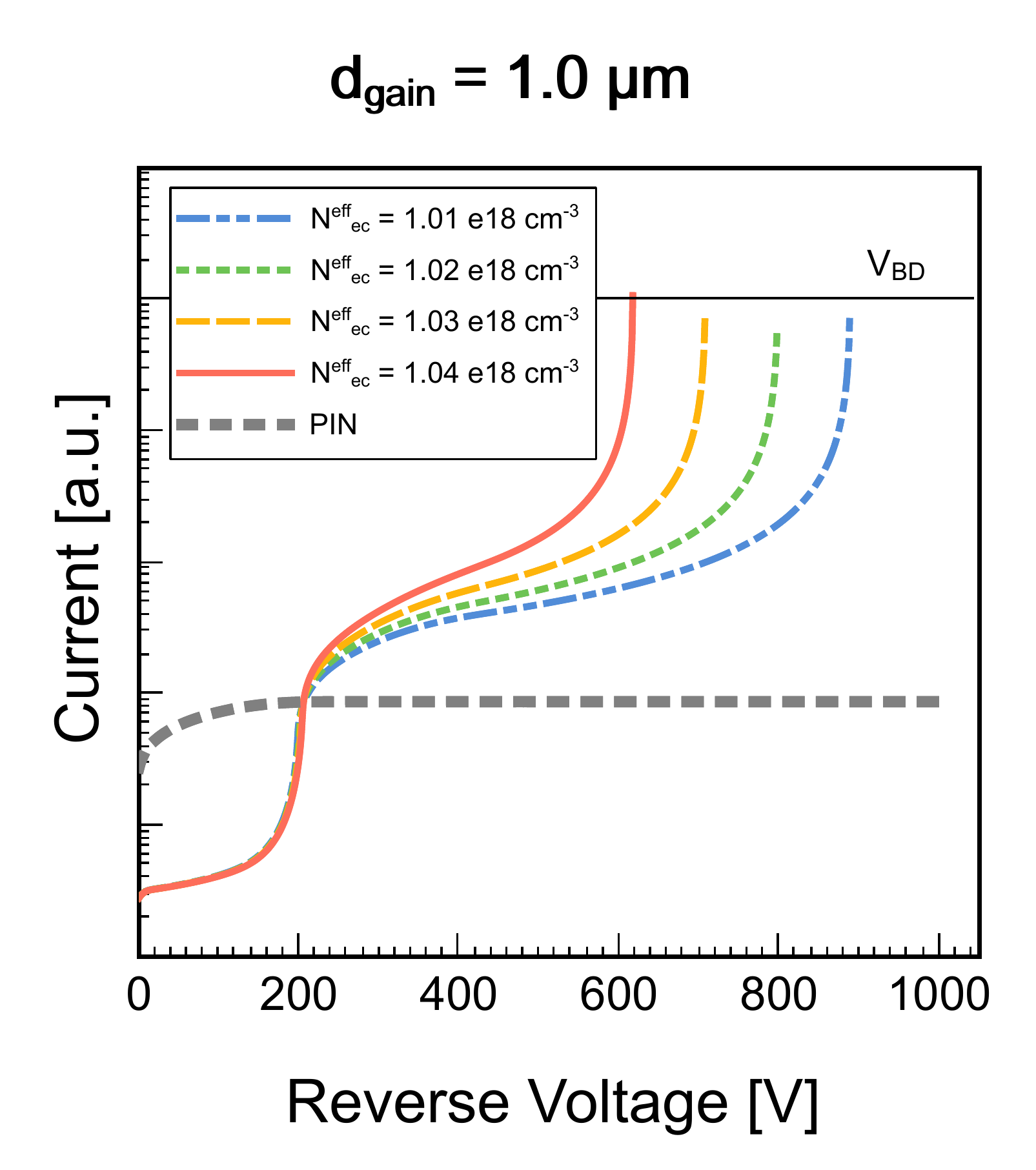}}
    \caption{Leakage current for simulation at 4 different electric field control levels of Trapezoid-Type 4H-SiC LGAD and PIN: (a)~$d_{gain}=0.5~\mu m$; (b)~$d_{gain}=1.0~\mu m$. The colored lines are Triangle-Type 4H-SiC LGAD. The dotted gray line is PIN.} \label{fig:Trapezoid_Type_IV}
\end{figure}

\begin{figure}[htb] 
    \centering 
    \subfigure[]{ \label{fig:Trapezoid_Type_CV_0.5um}
    \includegraphics[scale=0.3]{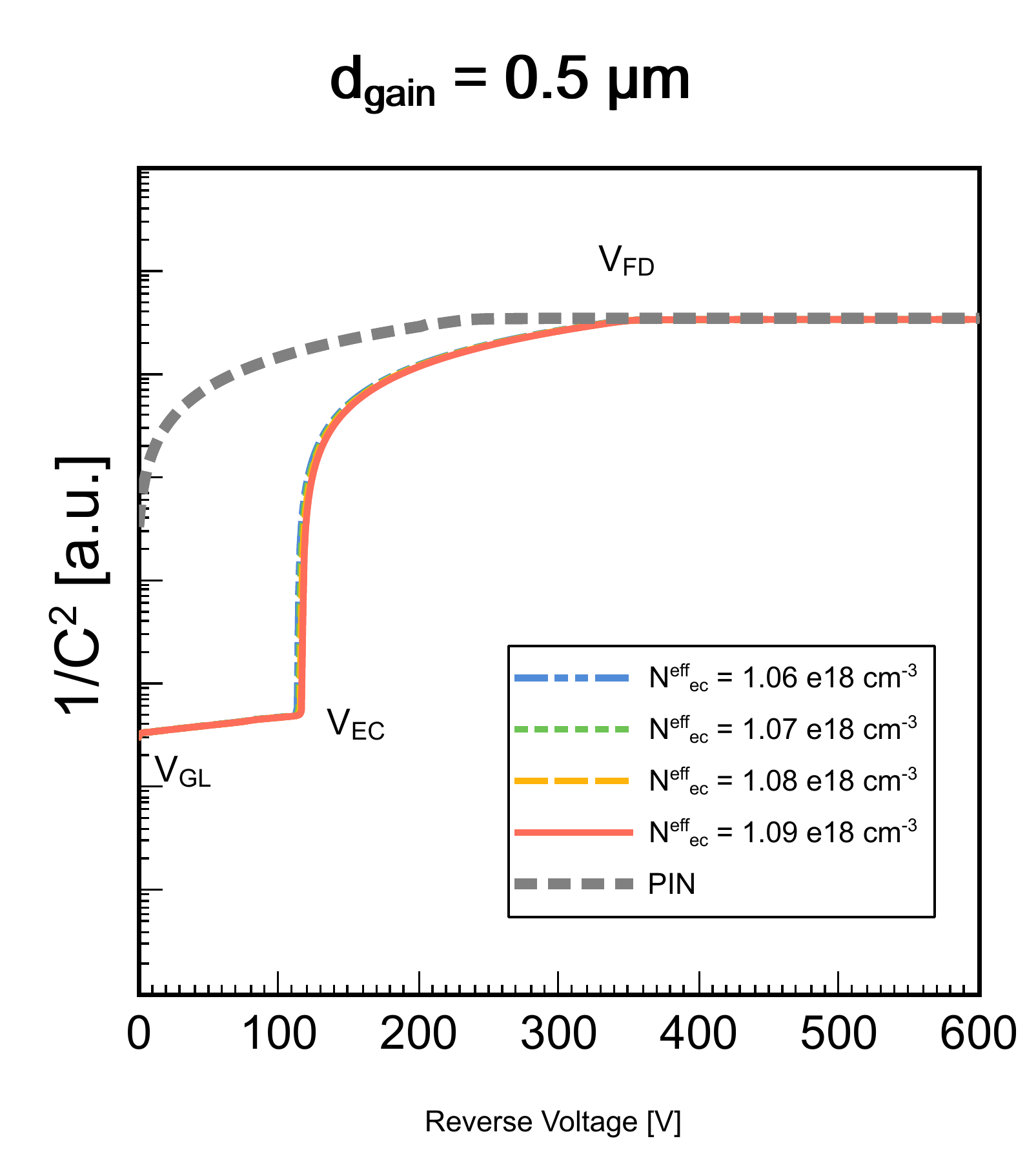}}
    \subfigure[]{ \label{fig:Trapezoid_Type_CV_1.0um}
    \includegraphics[scale=0.3]{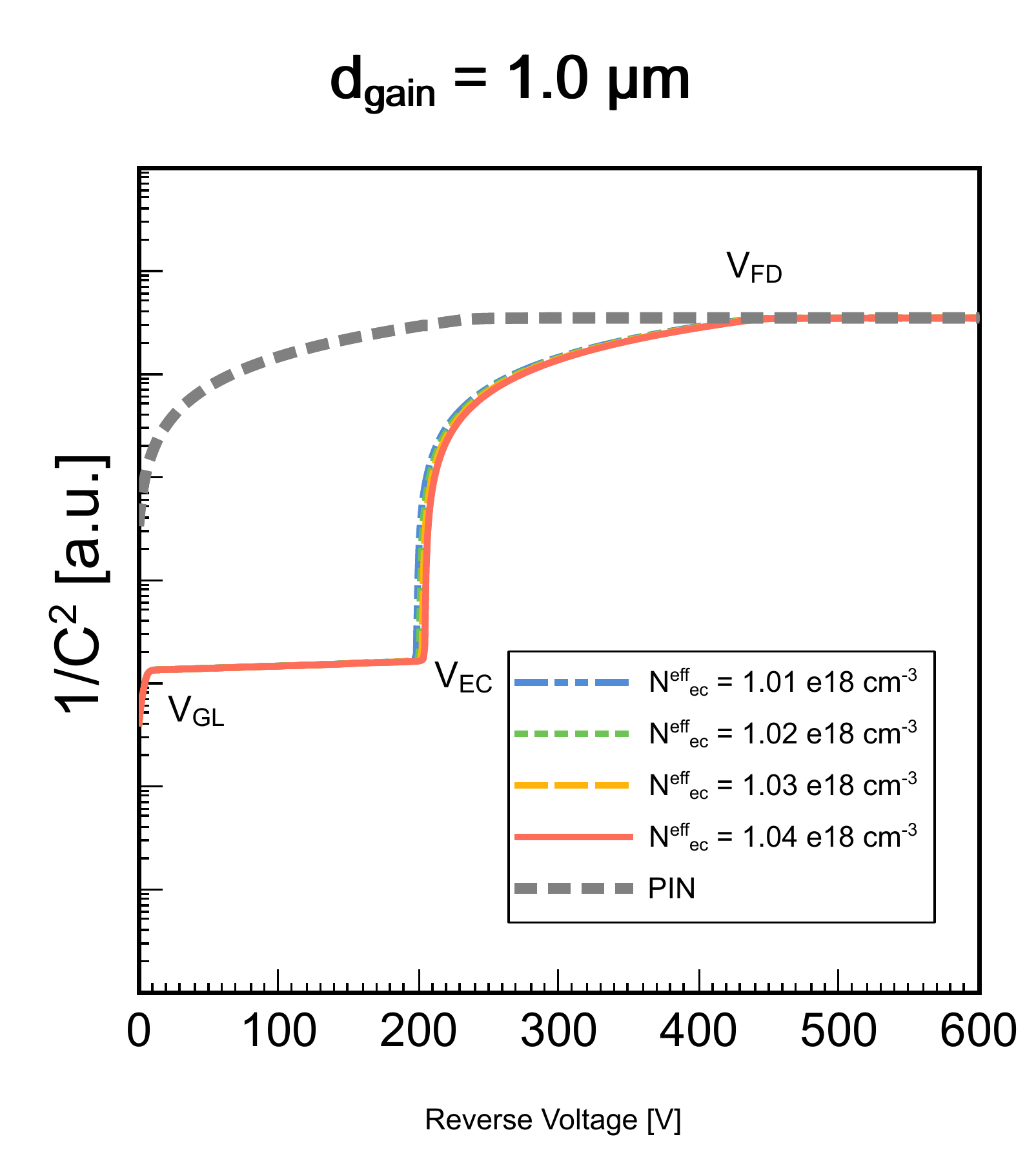}}
    \caption{1/C$^{2}$-V for simulation at 4 different electric field control layer doping levels of Trapezoid-Type 4H-SiC LGAD and PIN: (a)~$d_{gain}=0.5~\mu m$; (b)~$d_{gain}=1.0~\mu m$. The colored lines are Triangle-Type 4H-SiC LGAD. The dotted gray line is PIN.} \label{fig:Trapezoid_Type_CV}
\end{figure}

Similarly, the I-V and 1/C$^{2}$-V curves are simulated for Trapezoid-Type 4H-SiC LGAD shown in \figurename{~\ref{fig:Trapezoid_Type_IV}} and \figurename{~\ref{fig:Trapezoid_Type_CV}}. Only $d_{gain}=0.5~\mu m$ and $d_{gain}=1.0~\mu m$ are studied because the electric field layer is fixed to $0.1~\mu m$. Same with the Triangle-Type 4H-SiC LGAD, the breakdown voltages could also be restrained in 500~V$\sim$1000~V and the full depletion voltage $V_{FD}<$ 500~V in our designs. It should be noted that another sudden-changing voltage $V_{EC}$ is appeared in \figurename{~\ref{fig:Trapezoid_Type_CV_0.5um}} and \figurename{~\ref{fig:Trapezoid_Type_CV_1.0um}} due to electric field control layer is added.

\subsection{Simulation of gain for MIP}

\begin{figure*}[htb] \label{fig:MIPs_Ampl}
    \centering
    \subfigure[]{ \label{fig:MIPs_Ampl_Triangle}
    \includegraphics[scale=0.3]{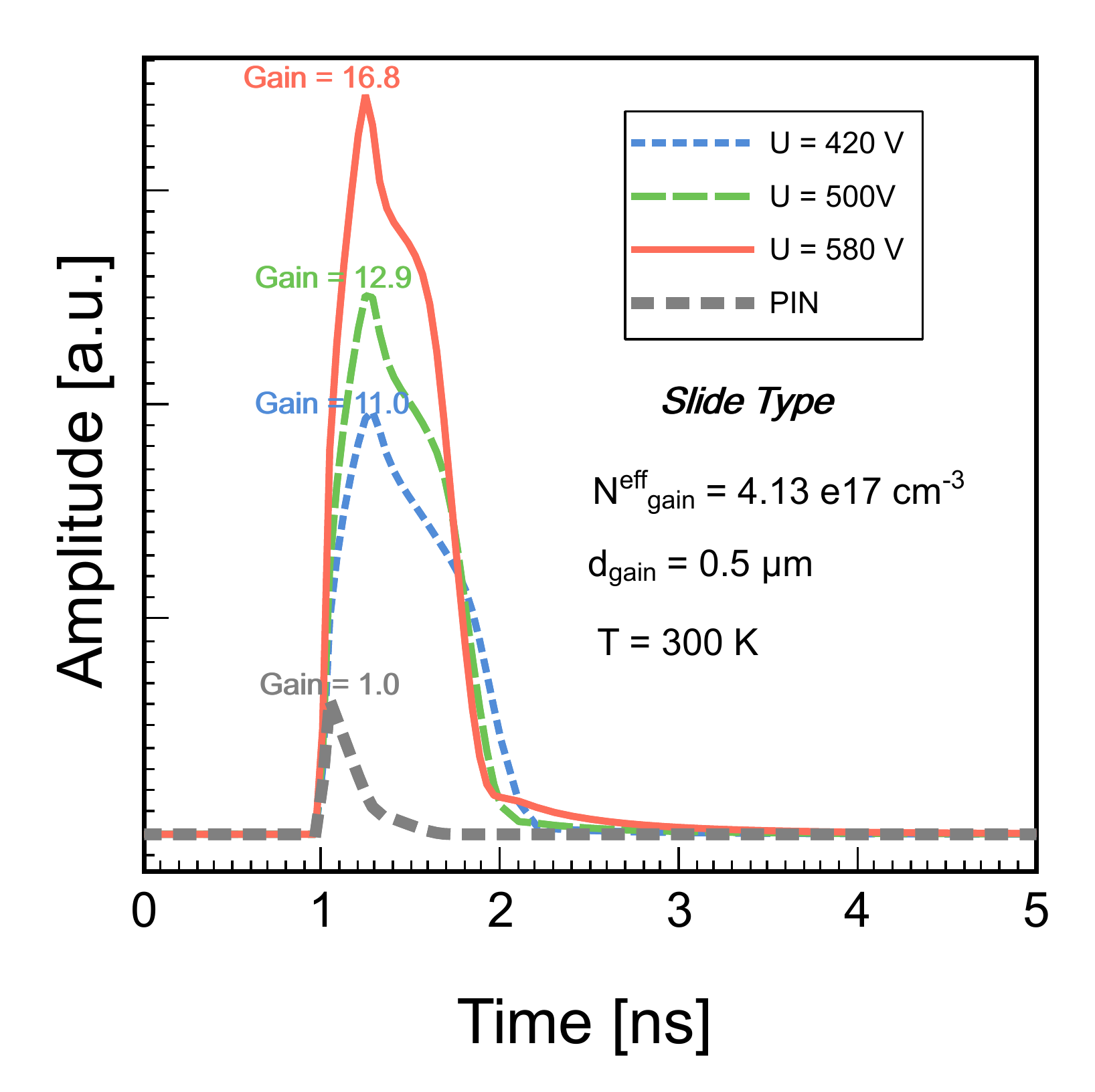}}  
    \subfigure[]{ \label{fig:MIPs_Ampl_Trapezoid}
    \includegraphics[scale=0.3]{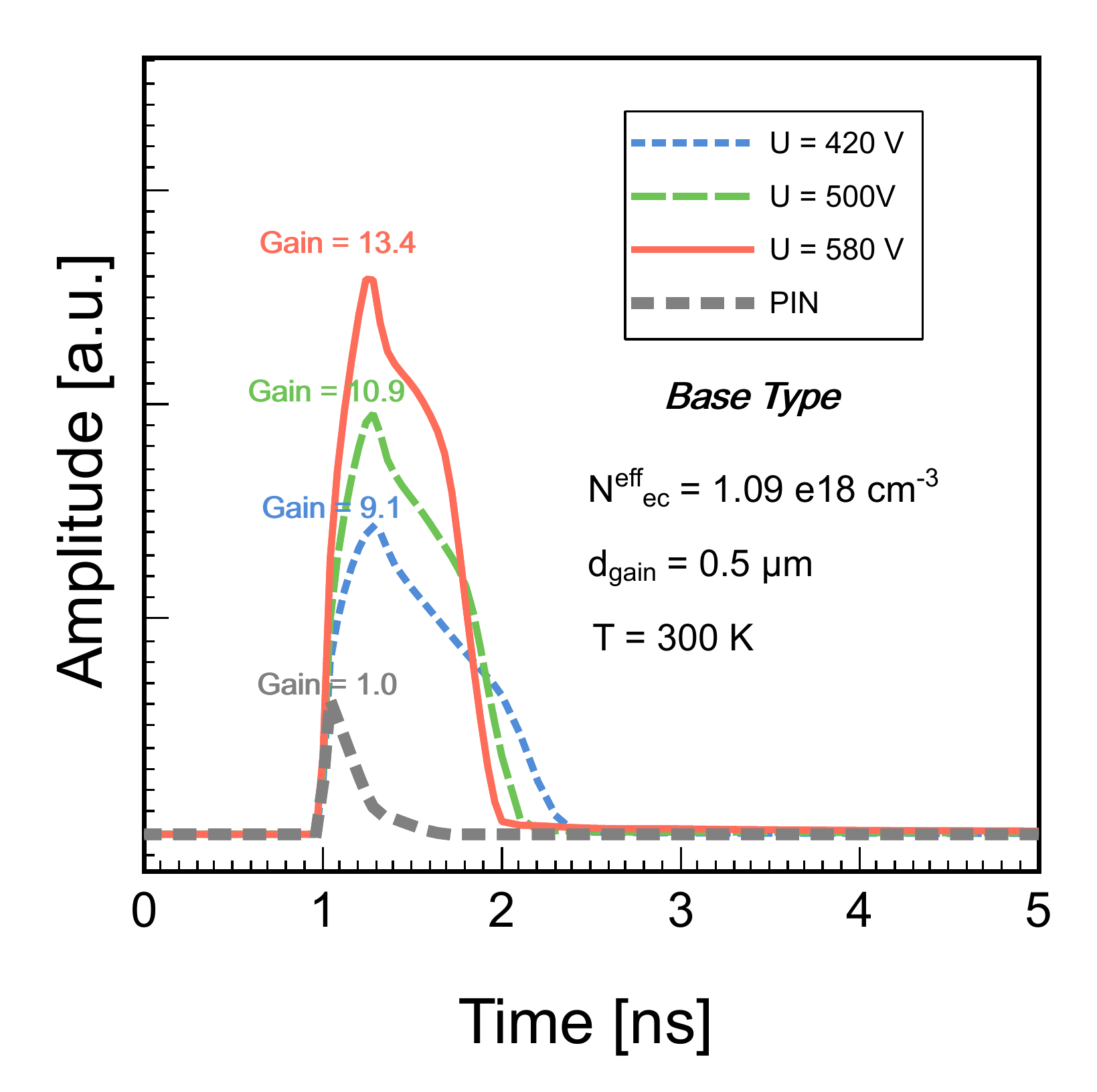}}
    \caption{(a)~the stimulated current by MIP in Triangle-Type 4H-SiC LGAD. The colored lines are Triangle-Type 4H-SiC LGAD under different operating voltages. The dotted gray line is PIN at U=500~V. (b)~the stimulated current by MIP in Trapezoid-Type 4H-SiC LGAD. The colored lines are Trapezoid-Type 4H-SiC LGAD under different operating voltages. The dotted gray line is PIN at U=500~V.}
\end{figure*}

\begin{figure*}[htb] \label{fig:MIPs_Ana}
    \centering
    \subfigure[]{ \label{fig:ElectricField}
    \includegraphics[scale=0.28]{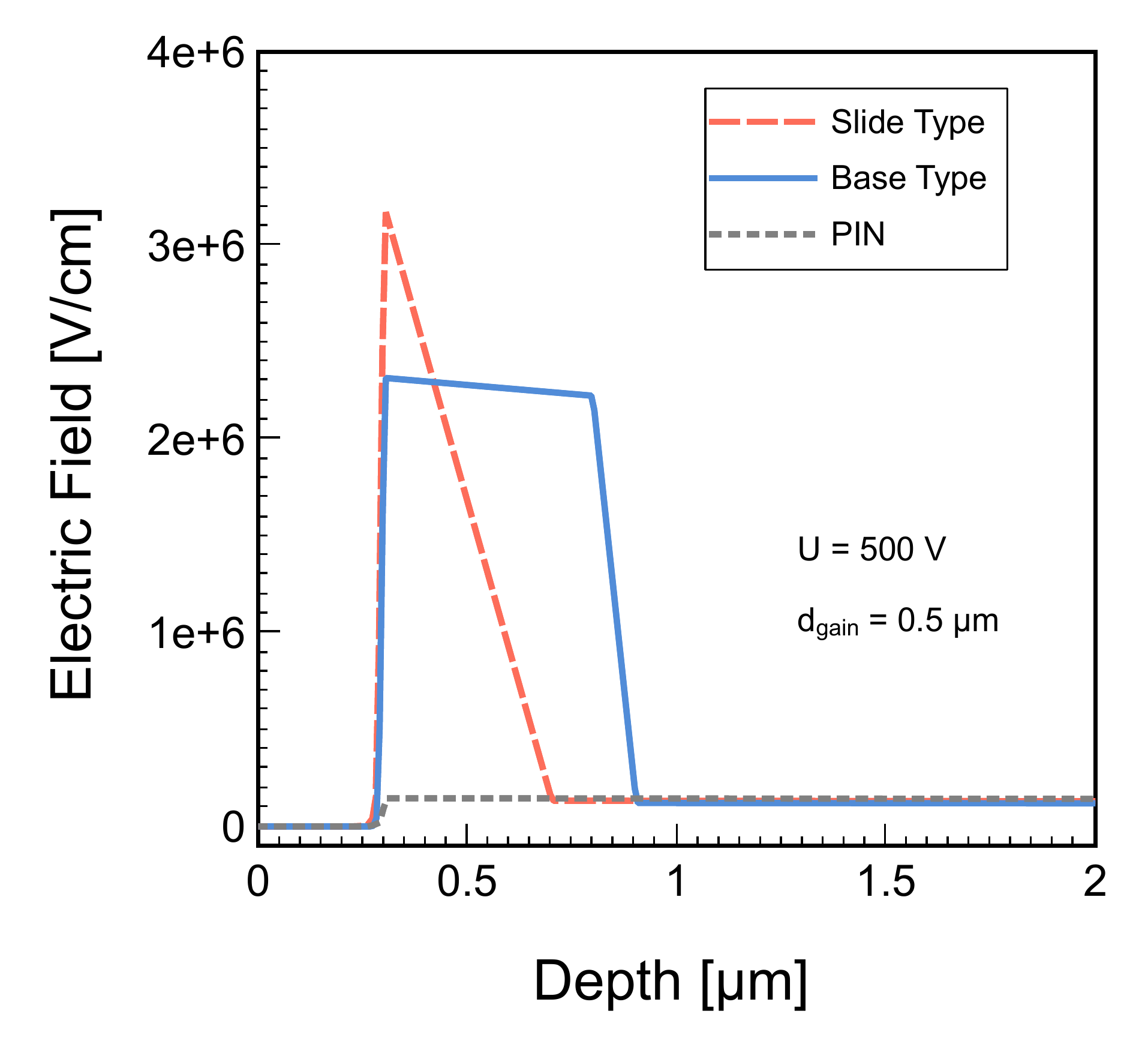}}
    \subfigure[]{ \label{fig:MIPs_Ampl_Slope}
    \includegraphics[scale=0.28]{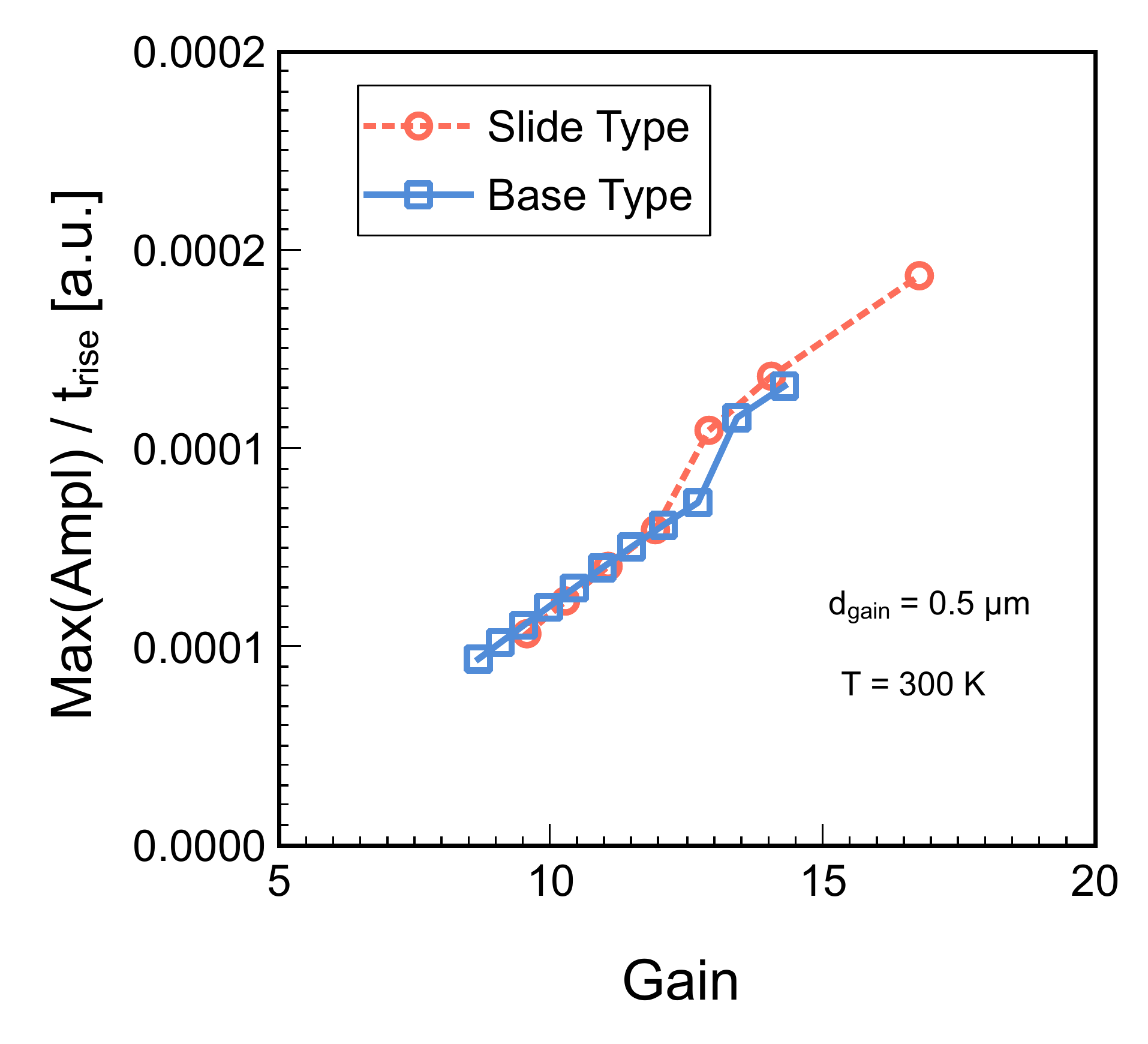}}
    \caption{(a)~the electric field near gain layer of Triangle-Type 4H-SiC LGAD, Trapezoid-Type 4H-SiC LGAD and PIN under the same operating voltage U=500~V. (b)~the slope of stimulated current by MIPs for Triangle-Type 4H-SiC LGAD and Trapezoid-Type 4H-SiC LGAD.}
\end{figure*}

To determine the gain factor when MIPs pass through the 4H-SiC LGAD, we use the heavy ion which has the same e-h pairs generation with MIP in 4H-SiC material to obtain the gain factor. The gain factor in this work is defined as:

\begin{equation}
    Gain = \frac{Q_{LGAD}}{Q_{PIN}} = \frac{\int_{0}^{T} i(t)_{LGAD}}{\int_{0}^{T}i(t)_{PIN}}
\end{equation}

where $Q$ is collected charges, $i(t)$ is stimulated current by MIP and $T$ is collection time. \figurename{~\ref{fig:MIPs_Ampl_Triangle}} and \figurename{~\ref{fig:MIPs_Ampl_Trapezoid}} demonstrate the stimulated current by MIPs under different operating voltages. Compare with the stimulated current of PIN, The current gain for both type of 4H-SiC LGAD is $>10$ when U=500~V. Under same operating voltage, the gain of Triangle-Type is higher than gain of Trapezoid-Type. It could be interpreted by using the profile of electric field shown in \figurename{~\ref{fig:ElectricField}}, where the higher electric field peak for Triangle-Type dominates the gain factor although the Trapezoid-Type has higher average electric field $\bar{E}_{gain}$. The electric field levels shown in \figurename{~\ref{fig:ElectricField}} agree with the predication in \figurename{~\ref{fig:impact_coff}}. Based on simulation, we determine the electric field is $\sim3.2\times10^{6} ~V/cm$ for Triangle-Type and $\sim2.3\times10^{6} ~V/cm$ for Trapezoid-Type to a 4H-SiC LGAD with $d_{gain}=0.5~\mu m$.

The time-response is described by the rising slope of stimulated current by MIP which is defined as $\frac{max(Ampl)}{t_{rise}}$, where the $max(Ampl)$ is the max value of stimulated current and $t_{rise}$ is rising time. \figurename{~\ref{fig:MIPs_Ampl_Slope}} demonstrates the direct proportion with gain and the rising slope. And it indicates both types of 4H-SiC LGAD have the same time response for the same gain. Combined with the previous analysis that the Triangle-Type has a higher gain than Trapezoid-Type, the Triangle-Type is better for the application of fast MIP detection.

\subsection{Efficiency of gain obtaining and stability}

\begin{figure*}[htb]
    \centering
    \subfigure[]{ \label{fig:Gain_RT}
    \includegraphics[scale=0.25]{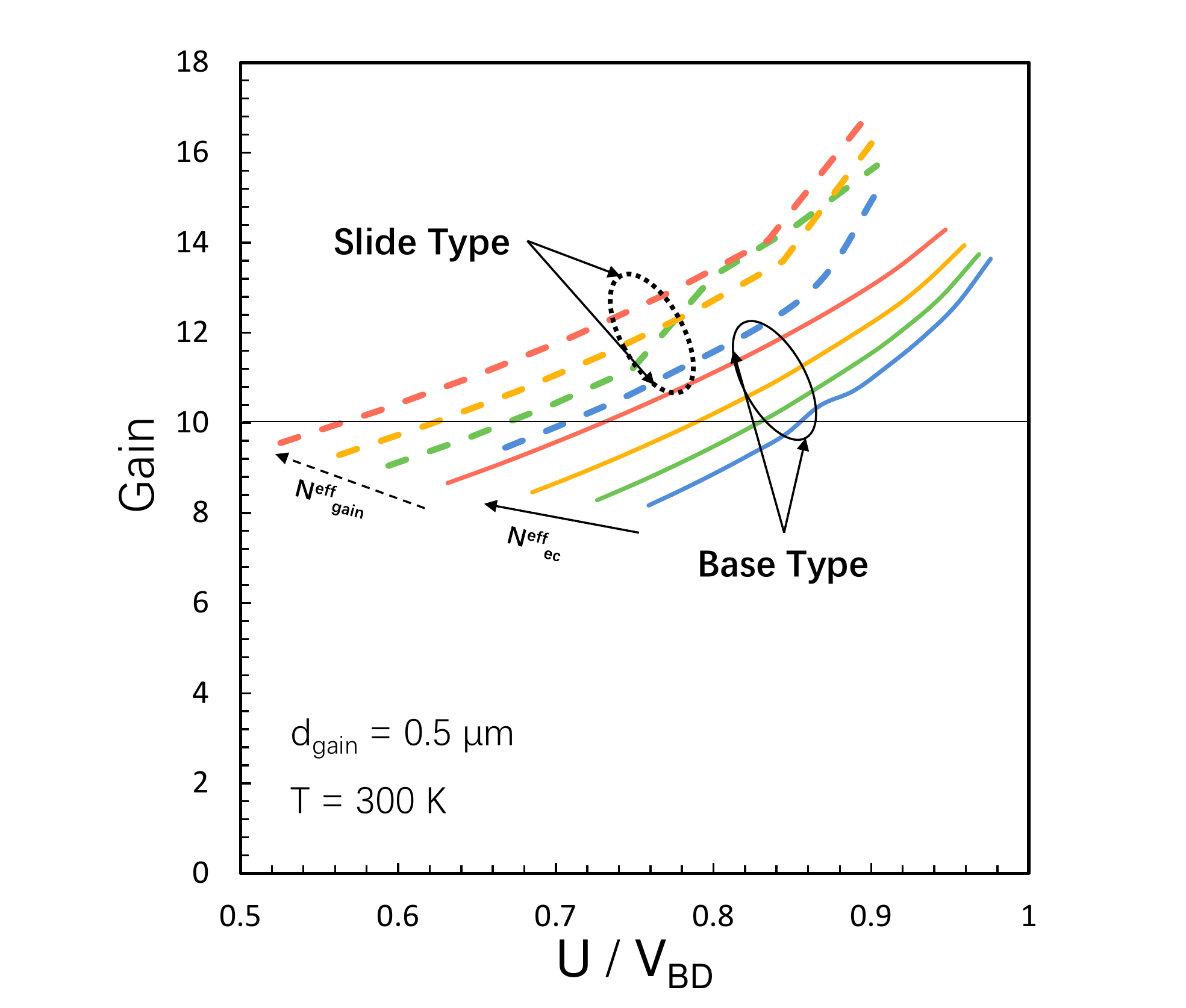}}  
    \subfigure[]{ \label{fig:Gain_d}
    \includegraphics[scale=0.26]{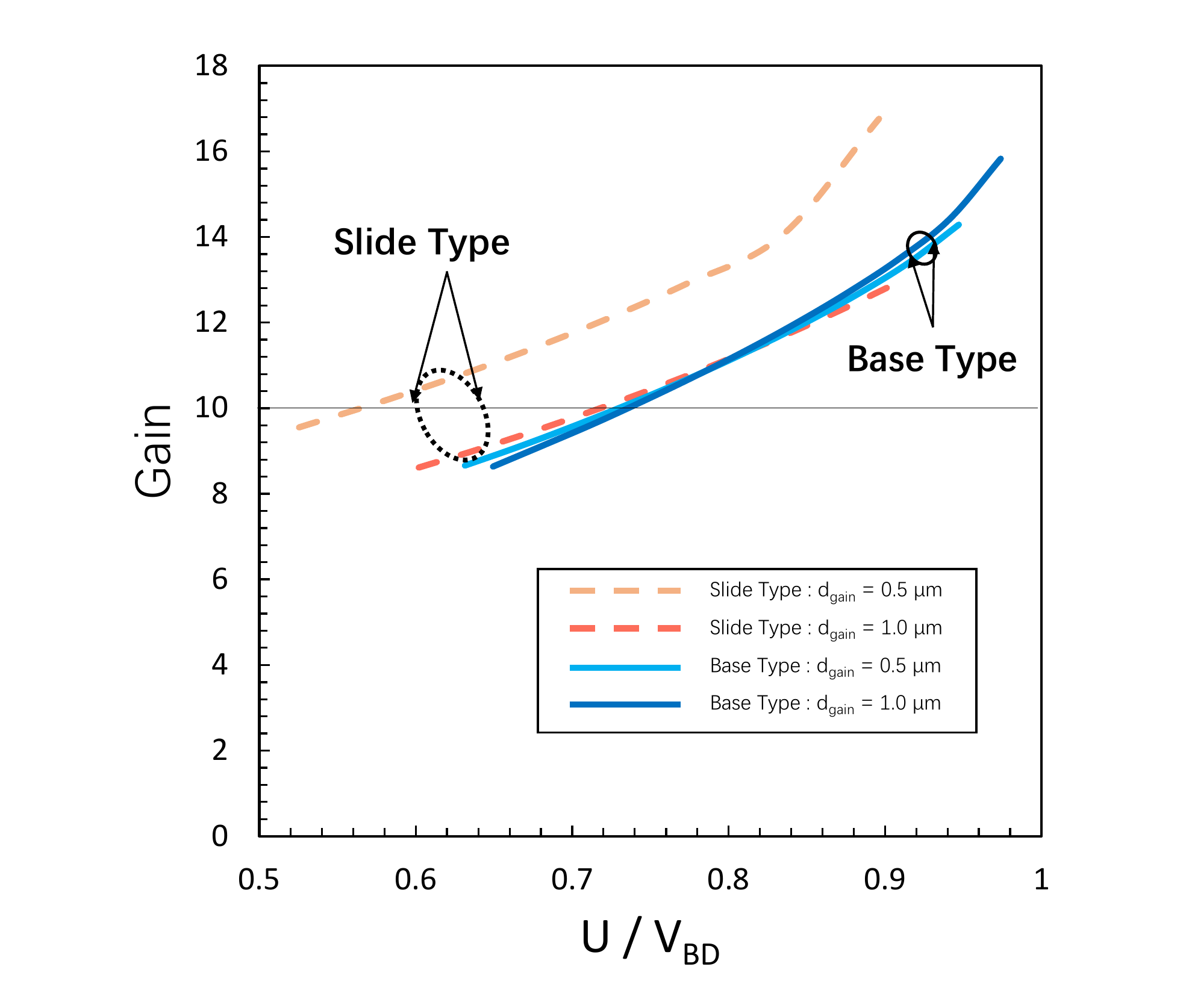}}
    \subfigure[]{ \label{fig:Gain_MT}
    \includegraphics[scale=0.26]{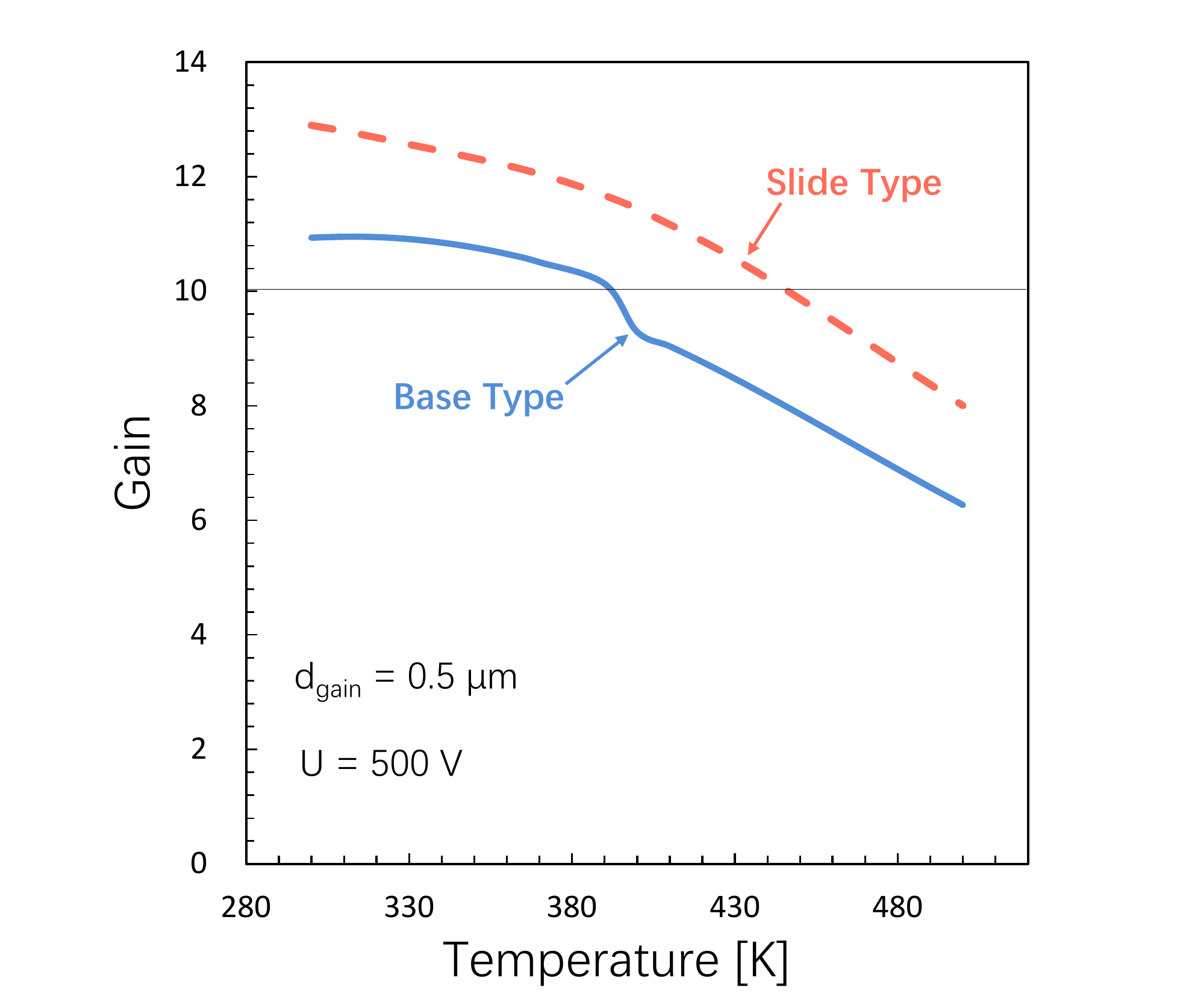}}
    \caption{(a)~the relations between gain and ${U}/{V_{BD}}$ for Triangle-Type 4H-SiC LGAD (dotted lines) and Trapezoid-Type 4H-SiC LGAD (solid lines). The different colored lines represents the different doping levels. (b)~the relations between gain and ${U}/{V_{BD}}$ for different thickness of gain layer. The dotted lineS are Triangle-Type and solid lines are Trapezoid-Type. (c)~the gain factor under operating voltage U=500~V at temperature from T=300~K to T=500~K.}
\end{figure*}

To compare the efficiency of gain obtaining for two types 4H-SiC LGAD, the relations between gain and the ratio ${U}/{V_{BD}}$ are studied. \figurename{~\ref{fig:Gain_RT}} shows the Triangle-Type 4H-SiC LGAD has higher efficiency of gain obtaining because it has a higher gain factor than Trapezoid-Type at the same ${U}/{V_{BD}}$ value. But it also indicates the Triangle-Type is easier to reach the breakdown threshold, as reflected in the rapidly increasing of gain when ${U}/{V_{BD}}>0.8$. Contrarily, all of Trapezoid-Type 4H-SiC LGADs behave more stable even if the ${U}/{V_{BD}}>0.9$.

Another piece of evidence that indicates the stability of the Trapezoid-Type 4H-SiC LGAD is shown in \figurename{~\ref{fig:Gain_d}}. When the thickness of the gain layer increases from $0.5~\mu m$ to $1.0~\mu m$, the $Gain-({U}/{V_{BD}})$ curves for Trapezoid-Type 4H-SiC LGAD are not changed that means the efficiency of gain obtaining do not rely on $d_{gain}$. For the Triangle-Type 4H-SiC LGAD, the gain factor decreases when the $d_{gain}$ increases at the same value of ${U}/{V_{BD}}$. It indicates the thinner gain layer has higher efficiency of gain obtaining but more easily breakdown. So the Trapezoid-Type 4H-SiC LGAD is better if consider its stability and robustness.

To investigate the temperature dependence and potential application in high temperature environment, the gain factors from T=300~K to T=500~K are simulated which are shown in \figurename{~\ref{fig:Gain_MT}}. Under the same operating voltage, the gain factor decreases with increasing of temperature. It is caused by the decreasing impact ionization coefficient depends on Hatakeyama model. The gain of both types are $>10$ when the temperature below 400K. 

\section{Summary} \label{sec:summary}

Based on the analytical analysis in Sec.~\ref{sec:design}, we provide the policy to determine the doping level and thickness of gain layer in 4H-SiC LGAD. In our work, $d_{gain}=~0.5~\mu m$ is adopted which has $50~\mu m$, $1\times10^{14}~cm^{-3}$ bulk layer. The two types of LGAD structures are compared by simulation where the Triangle-Type has ``triangle'' electric field and  Trapezoid-Type has ``trapezoid'' electric field. In the simulation, the gain $>10$ is achieved when the operating voltage U between $V_{FD}$ and $V_{BD}$. The Triangle-Type 4H-SiC LGAD is better for the application of fast MIP detection but also easily premature breakdown. The Triangle-Type 4H-SiC LGAD shows high stability and robustness but low efficiency of gain obtaining. It should be noticed that the simulation in this work is based on the specific configuration of physical parameters for 4H-SiC parameters. Different physical models  impact the final results. Especially for the impact ionization coefficient predicated by different models\cite{Hatakeyama_Impact_Model,Loh_Impact_Model,Niwa_SiC_Impact,SiC_Impact_High_T}, the difference could be some magnitudes in 4H-SiC in our studied temperature range. So the study of practical devices is necessary in the future. The preliminary results of Trapezoid-Type 4H-SiC LGAD fabricated by Nanjing University (NJU)\cite{NJU_SIC_LGAD_1,NJU_SIC_LGAD_2} have already verified the process technology. Further study and optimization of NJU 4H-SiC LGAD are on going. We believe the more interesting results could be obtained for practical devices in the future. And it will help us to understand the impact ionization in the thin 4H-SiC layer like 4H-SiC LGAD structure.


\section*{Acknowledgment}
This work has been supported by the National Natural Science Foundation of China (No. 11961141014), the State Key Laboratory of Particle Detection and Electronics (No. SKLPDE-ZZ-202218) under the CERN RD50 Collaboration framework.

\bibliographystyle{unsrt}
\bibliography{p18_sic_lgad_sim}

\end{document}